\documentclass[sigconf]{acmart} 

\AtBeginDocument{%
  \providecommand\BibTeX{{%
    \normalfont B\kern-0.5em{\scshape i\kern-0.25em b}\kern-0.8em\TeX}}}

\copyrightyear{2026}
\acmYear{2026}
\setcopyright{cc}
\setcctype{by}
\acmConference[IMX '26]{ACM International Conference on Interactive Media Experiences}{June 09--11, 2026}{Athlone, Ireland}
\acmBooktitle{ACM International Conference on Interactive Media Experiences (IMX '26), June 09--11, 2026, Athlone, Ireland}
\acmDOI{10.1145/3788851.3805017}
\acmISBN{979-8-4007-2448-0/2026/06}

\usepackage{packages}
\usepackage{notations}

\usepackage[commandnameprefix=ifneeded]{changes} 
\usepackage{xcolor} 

\newcommand{\addedcr}[2][]{#2}
\newcommand{\deletedcr}[2][]{}

\newcommand{\addedimx}[2][]{#2}
\newcommand{\deletedimx}[2][]{}
\newcommand{\replacedimx}[3][]{#2}

\renewcommand{\added}[2][]{#2}
\renewcommand{\deleted}[2][]{}
\renewcommand{\replaced}[3][]{#2}

\renewcommand{\chcomment}[2][]{}

\definechangesauthor[name=A1, color=green]{a1}
\definechangesauthor[name=A2, color=red]{a2}

\begin{document}
		\title[Gesture-Based Magnification versus Direct Access Reading]{Quantifying the Cost of Manual Navigation: A Comparison of Gesture-Based Magnification versus Direct Access Reading in Digital Layout-based Documents}

\author{Sebastián Gallardo}
\email{sebastian.gallardo-diaz@inria.fr}
\additionalaffiliation{%
  \institution{Demain Un Autre Jour} 
  \city{Toulouse}
  \country{France}
}
\author{Hui-Yin Wu}
\author{Dorian Mazauric}
\author{Pierre Kornprobst}
\email{pierre.kornprobst@inria.fr}
\affiliation{%
  \institution{Université Côte d'Azur, Inria}
  \city{Sophia-Antípolis}
  \country{France}
}

\author{Monica Di Meo}
\email{di-meo.m@chu-nice.fr}
\author{Stéphanie Baillif}
\email{baillif.s@chu-nice.fr}
\affiliation{%
  \institution{\addedcr{CHU Pasteur}} 
  \city{Nice}
  \country{France}
}

\author{Aurelie Calabrese}
\email{aurelie.calabrese@univ-amu.fr}
\affiliation{%
  \institution{Aix-Marseille Université, CNRS} 
  \city{Aix-Marseille}
  \country{France}
}

		\renewcommand{\shortauthors}{Gallardo, et al.}
\begin{abstract}
\addedimx{Understanding how diverse audiences engage with structured media is critical to ensure a consistent quality of experience. In this context, we quantify the behavioral and performance cost of manual navigation (e.g., pinch and zoom) versus direct structural access in layout-based digital documents. We specifically investigate newspaper reading when visual access to structural cues (headlines as entry points) is constrained.} \deletedimx{In particular, we investigate how readers engage with layout-based digital documents when visual access to structural cues (headlines as entry points) is constrained. Newspapers are a prime example, where headlines structure navigation across blocks of content.}  Participants completed two tasks—reading all headlines aloud and locating target articles—under two conditions: (1) original edition with gesture-based magnification (pan and zoom), \addedimx{which is the industry standard for digital documents}, and (2) large-print edition supporting direct-access reading. We collected performance measures (success ratio and completion time), behavioral integrity through reading path analysis, alongside perceived workload and preferences (NASA-TLX). Results from linear mixed-effects models show that the large-print condition yielded not only better performance than gesture-based magnification (18\% improvement in reading speed, 30\% improvement in speed to locate a target)\deletedimx{, and better preservation of the natural reading path}, \addedimx{but more importantly, restored the natural reading strategy that gesture-based magnification interaction disrupts}. Readers also reported lower workload and higher preference. These findings highlight the importance of developing automated methods for generating large-print editions, where layout adaptation complements font scaling to support accessibility \addedimx{and quality of experience}.

\end{abstract}

\begin{teaserfigure}
\centering
  \includegraphics[width=0.8\columnwidth]{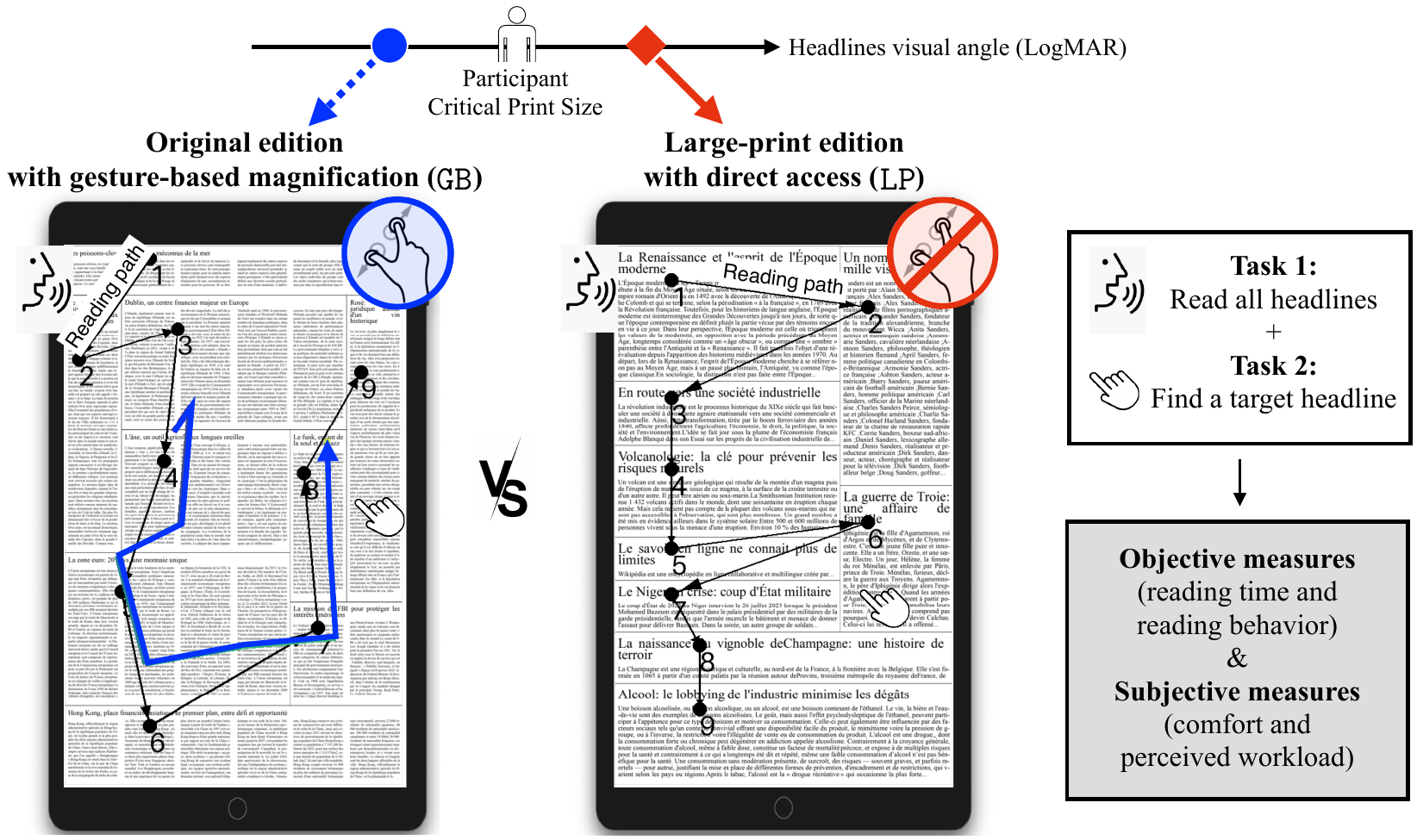}
  \caption{\label{fig:teaser} Experimental protocol for quantifying the behavioral and performance cost of digital newspaper consumption under visual constraints. The figure illustrates the two conditions used to evaluate the impact of manual navigation versus direct structural access when reading digital newspapers. In the gesture-based magnification condition (\condonenot), participants read the original layout, in which headlines were rendered below each individual's Critical Print Size (CPS), requiring pan-and-zoom magnification (high manual control, fragmented global view). In the large-print edition with direct access condition (\condtwonot), headlines were enlarged above CPS and directly legible without magnification (no manual control, direct structural awareness). 
  Across multiple trials, participants performed two tasks—reading all headlines aloud and locating target articles—while both objective performance and subjective responses were collected.
  }
\Description{The figure illustrates two ways of reading digital newspaper pages. On the left, the original edition is shown, where headlines are too small and require pan-and-zoom gestures. On the right, the same page is shown as a large-print edition, where headlines are enlarged and directly legible without magnification. Both versions contain multiple article blocks with headlines serving as entry points. Above these images, a linear axis indicates the Critical Print Size (CPS) of a participant, showing that in the original edition headline size falls below CPS, while in the large-print edition it falls above CPS. On the right-hand side, the figure also summarizes the experimental tasks—reading all headlines and locating a target headline—and the measures collected: objective measures (reading time, navigation behavior) and subjective measures (comfort, perceived workload).}
\end{teaserfigure}

\maketitle

\section{Introduction}
\label{sec:introduction}

Digital media reading spans a continuum from linear narratives to layout-based documents such as newspapers. In layout-based media (such as newspapers), information is distributed across multiple spatially distinct blocks. Readers must not only access text at a legible size but also maintain awareness of the document’s structure \addedimx{to ensure a consistent quality of experience.} This dual requirement—legibility plus structural awareness—distinguishes layout-based reading from simpler linear reading. Entry points such as headlines, subheadings, and visual cues guide attention and shape \replacedimx{digital newspaper consumption}{reading behavior} in ways that are unique to these documents~\cite{holmqvist_role_2005, zambarbieri_eye_2008, holmberg_eye_2004, eraslan_eye_2015}.

Effective navigation therefore hinges on visual access: the ability to perceive spatial organization and identify entry points. And so, navigation strategies break down when visual access is constrained and these cues can no longer be easily perceived. We refer to such situations as constrained visual access scenarios (\CVAS). In this study, we focus on two complementary \CVAS that both reduce the legibility of entry points: (1) device-related constraints, where limited screen size restricts visibility even for readers \NV\footnote{The World Health Organization (WHO) classifies visual impairment based on best-corrected visual acuity in the better eye, ranging from “Normal vision” (20/20) to moderate (down to 20/160), severe (down to 20/400), and profound low vision (down to 20/1000).} (e.g., on small phones) and (2) reader-related constraints, where reduced visual acuity limits access to layout structure even on larger displays (readers \LV). \footnotemark[\value{footnote}]. While these scenarios have been studied in the simpler context of plain text~\cite{atilgan2020}, their impact on digital newspaper, which have multi-block layouts remains under-explored.

These constraints have real-world consequences that extend beyond basic legibility. Because both types of \CVAS limit access to a page’s structure and entry points, they hinder reading, reduce exploratory behavior, and degrade the overall quality of  experience~\cite{Nielsen2011, Budiu2015, moran2016reading}. For readers \LV, the same loss of visual access further restricts access to newspapers, affecting not only information acquisition but also social participation, as news media play a key role in maintaining engagement with societal discourse~\cite{Hajek2023}. Ensuring that layout-based documents remain navigable under \CVAS is therefore essential for equitable access to information.

A common approach to support reading under \CVAS could be to simplify the content, for example, by linearizing articles into feed-like formats. While effective for legibility, these representations discard the spatial organization that fosters serendipity, engagement, and aesthetic appreciation in layout-based documents~\cite{ozretic_dosen_key_2018, hollander_e-reader_2011}.

When simplification is not desirable, the natural alternative is to magnify the original content. On touch devices, gesture-based magnification (namely \panzoom) is the standard solution. Its adoption is well-founded: manipulable interfaces increase users’ sense of control, predictability, and engagement~\cite{aljoudi, shneiderman_direct_1983, atata_evaluating_2025, priyadarshini_impact_2024}, including for readers with visual impairments~\cite{granquist_how_2018}. However, these benefits often come at the cost of global context~\cite{nav-patterns-2002, buring_zoomable}. While \panzoom preserves the underlying layout, it exposes it only through successive zoomed-in fragments. Readers must zoom in to read a headline, pan to follow the article, then pan again to locate the next entry point—or zoom out to regain a sense of the page. \addedimx{ Then, manual magnification does not just slow down the reader; it imposes a shift from the natural content consumption.} This fragmentation disrupts \replacedimx{the natural reading path}{reading flow}, a problem well documented in maps and linear documents~\cite{bowers-reading, tang_screen_2023, magnification-low-vision, nav-patterns-2002}. Yet, prior findings  cannot be assumed to generalize to layout-based documents. Newspapers feature multiple spatially distinct blocks, diverse entry points, and layered visual hierarchies~\cite{ozretic_dosen_key_2018}, creating navigation demands that differ from maps or linear text.

To address these limitations while maintaining both legibility and spatial structure, another option is to consider large-print layouts (\condtwo), inspired by historical initiatives such as the New York Times Large Type editions~\cite{nyt-large-type-1967}. These editions \addedimx{not only enlarge fonts (which might be not sufficient since layouts were not designed for this new font size~\cite{gallardo}), but also} reorganize layout blocks so that all entry points are directly legible without manual zooming. \addedimx{While this reorganization changes the absolute position of each element, it is specifically designed to preserve the 2D spatial paradigm of a newspaper. By maintaining a non-linear arrangement of multiple articles, the spatial qualities of the layout are preserved, ensuring the reader still engages with a 2D document rather than a simplified 1D linear representation.} Recent computational work~\cite{gallardo}, based on responsive design principles~\cite{responsive-design}, automates this process by increasing headline size and reducing the number of visible lines per block. By magnifying entry points while preserving spatial layout, \addedimx{this approach acts as a direct-access reading modality},  eliminating the need for gesture-based navigation, prioritizing global context at the expense of direct manipulation.

\replacedimx{Our study quantifies the behavioral and performance cost of these two modalities: }{Our study precisely examines this trade-off:} (1) whether the loss of global context undermines the advantages of manipulable interfaces, and (2) whether producing \condtwo layouts—designed to preserve global context but requiring effort or automation—is justified and for whom. To address this, we investigate how \condone and \condtwo layouts influence behavioral integrity, reading performance, and perceived workload under constrained visual access. We hypothesize that while \panzoom allows detailed inspection, it results in not only higher completion time than \condtwo (an anticipated mechanical consequence due to the elimination of manual navigation), but more importantly will disrupt the canonical reading strategy and increase the perceived workload in comparison to \condtwo. \addedimx{Our goal is to precisely quantify these costs through a multidimensional evaluation of performance, behavior and workload.}

To test these hypotheses, we conducted a controlled user study with participants having normal vision or low-vision, comparing performance (success ratio and completion time), reading behavior (reading paths), and page structure memorization (success ratio on finding specific headline) on text-only newspaper pages under two conditions (see Fig.~\ref{fig:teaser}):
\begin{itemize}
    \item [(\condonenot)] Reading the original edition with gesture-based magnification, required to access and read the headlines due to constrained visual access.
    \item [(\condtwonot)] Reading the large-print edition, with enlarged text and reorganized layout blocks (based on~\cite{gallardo}), ensuring all headlines are directly legible.
\end{itemize}
We evaluate both objective measures (success ratio, reading time, reading paths) and subjective measures (comfort and perceived workload) across two \CVAS reflecting different devices and readers' visual acuity: small-screen reading by participants \NV, and large-screen reading by participants \LV.

The remainder of the paper reviews related work, introduces our research questions and hypotheses, describes the experimental design, presents the results, and concludes with implications for the quality of experience in digital media consumption.

\section{Related work} 
\label{sec:related-work}
\subsection{Digital media reading}

\addedimx{Recent research in the interactive media experience community has begun to explore how different media-mixes, ranging from traditional articles to 360-degree video, affect user experience in immersive journalism. For instance, Bujić et al.~\cite{bujic-imx2023} compared 360-video across devices against a 'monitor-article' condition, finding that while immersive formats may increase involvement, traditional text-based articles can offer lower levels of distraction, which is critical in contemporary news consumption. This highlights the relevance of text-based media on digital platforms.}

Within this landscape, prior research has shown that mobile devices have reshaped reading practices and have changed readers' habits~\cite{readinghabits2016, reading-periodicals2021, mobilereading2015}.
At the same time, reading on mobile remains more demanding than on larger displays.
Budiu~\cite{Budiu2015} identified inherent constraints—such as limited screen size, frequent interruptions, single-window views, and unstable connectivity—that require careful design considerations to support sustained reading.

In particular, Nielsen~\cite{Nielsen2011} reported that comprehension of complex content on small displays is only about 48\% of that on desktop monitors. This gap is explained by two main factors: mobile readers see less information at once, reducing context and thus understanding, and they need to navigate through the document, which diverts attention. Moran~\cite{moran2016reading} further shows that while short, simple texts are read with similar comprehension across devices, reading speed drops significantly for longer or denser passages, highlighting the need for layout optimization and interaction strategies to improve mobile reading performance.

These challenges become even more acute with structured, layout-driven documents. These structured documents combine dense text with visual design and spatial organization to convey meaning. \added[comment="2. Why LP?"]{Due to their distinct design elements, findings derived from linear reading cannot be generalized to layout-driven documents~\cite{ozretic_dosen_key_2018,hollander_e-reader_2011}.} They pose distinctive challenges due to their complex structures, often organized around visual entry points~\cite{holsanova_entry_2006}. 

\added{It is worth noting that, in theory, web-responsive adaptation principles could address the general challenge of reading complex documents on small screens. However, preserving spatial semantics is often at odds with standard web-adaptation paradigms. Techniques such as responsive design handle small displays primarily through linear re-flow mechanisms~\cite{reflow, marcotte_responsive_2010}, collapsing multi-column grids into a single vertical stack defined by the DOM order. While this method preserves sequential reading order and eliminates bi-directional scrolling, it sacrifices the 2D spatial paradigm that defines newspapers and other layout-driven documents~\cite{chesham_master_2003, chiou_2024}.}

\added{As a result, for documents whose navigation, visual hierarchy, and interpretation rely on spatial organization~\cite{holmqvist_role_2005, ohara_comparison_1997}, standard responsive design—and similarly summary-based or linearized representations—cannot serve as viable alternatives: they remove visual entry points, suppress layout cues, and fundamentally transform the nature of the reading task.
}

This limitation motivates evaluating large-print layouts as an alternative that preserves spatial organization while restoring entry-point legibility, allowing us to examine \replacedimx{the behavioral and performance cost of manipulable navigation (\condonenot) compared to direct structural access (\condtwonot).}{the trade-off between manipulable control (\condonenot) and global awareness (\condtwonot)}.

\deleted{Addressing navigation in these contexts resonates with GUI research. For instance, Shneiderman’s ``Visual Information Seeking Mantra'' \cite{shneiderman1996} outlines a progression from global overview to local zooming and filtering, and finally to details-on-demand. This principle directly connects to the magnification strategies explored in our work, bridging document design and interaction techniques.}

\subsection{Situational visual impairments (SVIs) in mobile reading}

Situational visual impairments~\cite{svi-challenges} (SVIs) arise when reading is disrupted by device limitations, such as small font or display size and low resolution, or by environmental factors, including luminosity and ambient noise. Tigwell et al.~\cite{tigwell-2018} investigated how SVIs like glare or movement affect mobile content usability, highlighting that designers often lack specialized tools and guidelines to address these challenges. They proposed preliminary recommendations and emphasized the need for improved design support through better resources and educational frameworks. Building on this, another study~\cite{svi-bright-env} specifically examined the impact of bright outdoor lighting on screen readability. Their results indicate that, while ambient brightness reduces readability, the intrinsic brightness and contrast of content play an even larger role, suggesting that design interventions should prioritize optimizing content contrast over solely compensating for glare.

In parallel, technical solutions have been proposed to mitigate SVIs. \emph{SituFont}~\cite{yue2024situfont} is a dynamic font adaptation system that adjusts attributes such as size, weight, and spacing based on real-time sensor inputs and human-in-the-loop feedback. 
The authors validated their approach through comparative evaluations across eight simulated SVI scenarios, considering factors such as environment (indoor vs. outdoor), user mobility, and luminosity. Their findings show that SituFont significantly improves reading efficiency and reduces cognitive and physical workload compared to manual adjustments, highlighting the potential of adaptive font systems for supporting mobile reading under situational constraints. 

While these studies provide valuable insights and technical solutions for reading under SVIs, they primarily focus on text consumption. Our work extends this line of research by investigating structured digital layout-based documents \replacedimx{under constrained visual access scenarios (\CVAS), examining the specific behavioral and performance cost of magnification strategies.}{, where reading is affected under constrained visual access scenarios (\CVAS).} In particular, we examine SVIs related to small headline font sizes and the need for magnification.

\subsection{Reading with low vision}

\addedimx{Despite the importance of inclusive design, users with disabilities have historically been underrepresented within the IMX community. A survey of 17 years of research~\cite{vatavu-imx2021} revealed that only 4.23\% of papers addressed accessibility, with a specific call for more empirical studies involving people with disabilities. This gap motivates our work, which evaluates reading behavior across two distinct scenarios of visual limitation. }

In this context, digital devices provide new opportunities for readers \LV to access information through adaptable text formats and presentation options~\cite{reading-digital-legge}. A growing body of research has examined how these readers interact with digital reading environments, including studies that simulate \CVAS to better understand performance under magnification and limited visual access. In this topic, a key contribution by Atilgan et al.~\cite{atilgan2020} provides a unified framework to evaluate how print size and display size impact reading speed for both small-display readers and individuals with low vision using magnified text. Their results show that limitations in font and display size can prevent some readers from maximizing reading performance. 

Several studies have examined display configurations and strategies for magnified reading. Xiong et al.~\cite{xiong_digital_2022} proposed guidelines emphasizing the importance of exceeding the critical print size (CPS) and maintaining at least 13 characters per line to support fluent reading. Granquist et al.~\cite{granquist_how_2018} asked participants with low vision to adjust text for comfortable reading, finding that most relied on enlarging text, and to a lesser extent on reducing viewing distance. Beckmann et al.~\cite{beckmann_psychophysics_1996} studied manual navigation with CCTVs (stand-mounted video magnifiers) and reported that reading speed decreases unless the viewing window is sufficiently wide (>10 characters), consistent with the threshold identified by Xiong et al.~\cite{xiong_digital_2022}. Magnification and contrast enhancement were also shown to improve reading speed in simulated low-vision environments, with magnification generally having a more substantial effect than contrast~\cite{magnification-low-vision}.

Other work has examined navigation and magnification strategies in more detail. Bowers et al.\cite{bowers-reading} analyzed reading with optical magnifiers, highlighting challenges such as line retrace that need to be addressed in future research. Tang et al.~\cite{tang_screen_2023} compared two types of screen magnification on modern devices: full-screen magnification (analogous to pan-and-zoom) and lens magnification (allowing users to see a larger portion of the un-magnified screen for spatial reference). They identified trade-offs in performance and usability, showing that lens mode led to more consistent and uniform mouse movements, whereas full mode caused longer and more frequent pauses. Similarly, Aguilar and Castet~\cite{aguilar-castet2017} developed a gaze-controlled system that magnifies a portion of text while maintaining a global view, which participants found more comfortable than traditional CCTVs.

Together, these studies highlight the impact of magnification on low-vision reading. However, most prior work focuses on plain, unstructured text, analyzing reading speed, errors, and navigation patterns within a fixed set of lines. In contrast, our study investigates structured layout-based documents, such as newspapers, where reading involves navigating not only text but also the spatial layout. We examine how visual access limitations affect document reading for both individuals with \NV and those \LV. 
\replacedimx{In particular, we examine the behavioral and performance cost of gesture-based magnification (\condonenot) compared to the direct structural access of large-print layouts (\condtwonot), quantifying how each strategy impacts the reading process. }{In particular, we further study the trade-off between the sense of control provided by gesture-based magnification and the global awareness of large-print layouts}


\section{Research Questions and Hypotheses}
\label{section:hypo}
We will hereafter use the abbreviations \condonenot (original edition, gesture-based magnification) and \condtwonot (large-print, direct access) to refer to the two experimental conditions.
Our study addresses the following research questions:

\begin{itemize}
    \item[RQ1] How do the two conditions differ in terms of reading behavior on a digital newspaper page?
    \item[RQ2] How do the two conditions affect participants’ ability to form an accurate mental representation of the page layout?
    \item[RQ3] Which condition is preferred by participants, and which one leads to a lower perceived workload?
\end{itemize}

From these research questions, we derive the following hypotheses:

\begin{itemize}
    \item[\hypperformance] Participants in \condtwonot will show shorter reading times and faster transitions between articles than in \condonenot.
    \item[\hyppath] \condtwonot maintains a reading strategy closer to the natural reading path of the original page than \condonenot.
    \item[\hypfindingperformance] In \condtwonot, participants will locate target headlines more quickly than in \condonenot.
    \item[\hypfindingpositions] In both conditions, the position of the target article will significantly influence finding time.
    \item[\hyppreference] Participants will prefer \condtwonot over \condonenot, reporting lower cognitive load and greater comfort.
\end{itemize}

To answer these research questions and test our hypotheses, we designed two tasks and a controlled scenario, described in the next section.

\section{Methods} 
\label{sec:method}

\subsection{Overview of the study}
\label{subsection:overview}

\subsubsection{General procedure}
We have conducted a \addedcr{2x2 split-plot design~\cite{lazar-hci-2017} study to quantify the behavioral and performance cost of digital newspaper documents under \CVAS.} We analyze how these individuals (readers with normal vision on small displays and readers with low vision even in larger displays) consume these digital documents.

\addedcr{Approval for this research was granted by Inria’s Operational Committee for the Assessment of Legal and Ethical Risks.}

To create controlled yet realistic reading conditions, we designed text-only newspaper pages, each composed of multiple blocks containing one headline and a body text (\secref{sec:newspaperpages}). This approach provides two key advantages:
\begin{itemize}
    \item Controlled entry points: Headlines serve as the sole entry points, ensuring that participants’ reading path relies on a textual structure that can be controlled and standardized. This also allows us to track the sequence in which participants access articles through aloud reading, a practical alternative to eye-tracking, which is difficult to implement with participants \LV.
    \item Reduced cross-trial variability: By avoiding graphical elements (e.g., advertisements, photographs, or diagrams), which are harder to standardize, we minimize potential confounds across trials and conditions, ensuring comparability of reading behavior data.
\end{itemize}

The experiment consisted of two tasks \addedimx{designed as controlled proxies of real-world newspaper consumption strategies, allowing for systematic observation and quantification of reading behavior:}

\begin{itemize}
\item {\bf Task 1 (structural orientation):  read all headlines.} \\\addedimx{This task emulates the initial scanning phase of newspaper consumption.} Participants read aloud all headlines on a page, as fast but intelligibly as possible, in any order they wished. Reading aloud allowed us to unambiguously capture the order in which headlines were read,  \added[comment="4. reading aloud."]{and is both a well-established method in reading research~\cite{wang_understanding_2023, wang_gazeprompt_2024} and clinical assessments (e.g., MNREAD).} This task addressed \hypperformance and \hyppath. There was no time limit. Participants decided themselves when they had finished, even if some headlines were missed.

\item {\bf Task 2 (targeted information retrieval): find a target headline.} \\ \addedimx{This task tests the user's mental representation of the page and their ability to retrieve specific information.} Participants were asked to locate a specific headline containing a given keyword, to probe their mental representation of page structure. This task tested the hypothesis \hypfindingperformance. To examine \hypfindingpositions, we divided the page into three vertical zones (top, middle, bottom) and selected two targets per zone. The order was randomized for each participant, and the zone division was never disclosed. Each search trial was limited to one minute.
\end{itemize}

Both tasks were performed under \condonenot and \condtwonot. A critical design feature of our protocol was to ensure that participants had to rely on \panzoom in \condonenot but could comfortably read headlines in \condtwonot. To achieve this, we estimated each participant’s Critical Print Size (CPS)~\cite{legge2006}, defined as the smallest print size allowing readers to maintain maximum reading speed, \added{and expressed in logMAR units (logarithm of the Minimum Angle of Resolution), which quantify text size in terms of visual angle rather than physical dimensions, which is crucial because it jointly depends on both print size in millimeters and viewing distance.} Headlines in \condonenot are below each participant’s CPS, forcing gesture-based magnification, while in \condtwonot they were \added{manually} set above CPS, enabling comfortable direct reading. 

\added[comment="3. Eye tracking."]{As a note, we also considered integrating eye-tracking to capture complementary gaze data. However, we opted against it due to specific challenges for this work. First, in this context, fixations mix cognitive inspection with smooth pursuit driven by viewport motion, and gaze positions must be continuously remapped to a moving document coordinate system, which limits interpretability~\cite{heo_reading_2024, jacob_eye_2003, valsecchi_saccadic_2013}. Second, there is a limitation to obtain a precise calibration with participants with low vision, since precise calibration requires an intact fovea which is usually not the case~\cite{wang_understanding_2023, heo_reading_2024}. Although technically feasible thanks to workarounds that mitigate the underlying issues, these constraints would have reduced the validity of the resulting metrics for our specific goal, which was to characterize reading strategies at a macroscopic level, not fine-grained oculomotor behavior. }

\begin{figure}[htbp]
    \centering
    \includegraphics[width=\widthForOneImage]{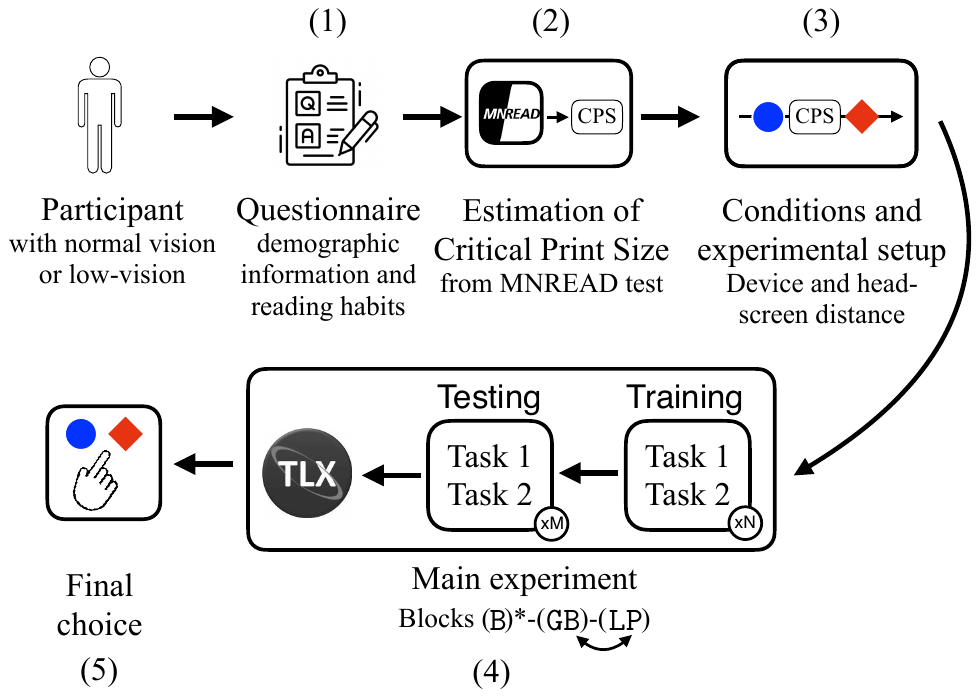}
    \caption{\label{fig:protocole-overview}
    Overview of the study. This figure illustrates the workflow of our experimental protocol, organized into five main steps: (1) questionnaires, (2) estimation of Critical Print Size (CPS), (3) conditions and experimental setup, (4) the main experiment, and (5) a final preference choice. The main experiment consisted of successive blocks in the two conditions (\condonenot and \condtwonot). \deleted{Participants \NV started with a baseline condition \condzeronot, after which all participants completed \condonenot and \condtwonot in counterbalanced order.} \added{Some adjustments were made depending on the participant scenario.} Each block included a training phase followed by test phases, where Tasks~1 and 2 were repeated across trials, and ended with the NASA-TLX questionnaire. The overall protocol lasted less than an hour, with the main experiment capped at 30 minutes to avoid excessive fatigue for participants with low vision.}
    \Description{The figure presents the experimental workflow in five steps: 
  (1) questionnaires, (2) Critical Print Size estimation, 
  (3) assignment to conditions and setup, 
  (4) the main experiment consisting of training and test blocks under \condonenot and \condtwonot conditions in random order, and 
  (5) post-experiment measures including NASA-TLX and final preference.}
\end{figure}

Following these principles, the study protocol unfolded in five steps, summarized in \figref{fig:protocole-overview}. 
\begin{enumerate}
    \item Participants first completed a short questionnaire about gender, age group, and reading habits (general and news-specific). 
     \item We ran a standard MNREAD test~\cite{legge2006} to estimate each participant’s Critical Print Size (CPS). 
    \item Based on the CPS, we defined how the two conditions would be instantiated (see \secref{sec:condition-definition}), including the choice of device (small-screen phone for participants with normal or corrected-to-normal vision, large-screen tablet for participants with low vision) and the viewing distance to be maintained during the experiment.
    \item After a detailed explanation of the protocol, participants proceeded to the main experiment. Each block corresponded to one condition, with the order counterbalanced across participants.
    \begin{itemize}
        \item To familiarize themselves with the tasks, participants first completed as many practice trials as needed. We used three fixed layouts (same order for everyone), and cycled through them if more than three trials were required. In practice, one or two trials were sufficient. 
        \item Participants then completed \deleted{six} $M$ test trials per condition. \added{Number of trials $M$ depending on the participant scenario (see Sections~\ref{sec:protocole-adjustements-nv} and \ref{sec:protocole-adjustements-lv})}. Layouts were randomized, and content was randomly assigned to conditions (see Section~\ref{sec:newspaperpages})
        \item After each block, we measured perceived workload using the NASA-TLX questionnaire~\cite{hart_nasa-task_2006}, \replaced{using a 7-point likert scale}{with ratings on a 0 (low)–6 (extreme) scale.}
    \end{itemize}
    \item Finally, participants indicated their overall preference between \condonenot and \condtwonot (or equal preference).
\end{enumerate}
This protocol served as the common foundation of our study. However, some adjustments were necessary depending on the participant's scenario.
\subsubsection{Protocol adjustments: participants \NV}
\label{sec:protocole-adjustements-nv}
\added{First, after pilot studies to study participants' fatigue, we decided to fix $M=6$ testing trials for participants \NV.}

\added{Second, we included an additional baseline condition (\condzeronot), applied before \condonenot and \condtwonot, with the same number of trials ($N$ for practice and $M=6$ for testing).}

\condzeronot simulates the experience of reading a physical newspaper, corresponding to unconstrained reading. In this condition, the original edition was displayed on a large-screen tablet, with no magnification gestures, as headlines were already comfortably legible (approximately $0.3$ logMAR at 40\,cm). Note that this baseline was not feasible for participants \LV, since the original edition would remain unreadable without magnification, even on a tablet. Data from \condzeronot were used to define a natural reading path, which we call reference paths, and these were compared to the paths observed in \condonenot and \condtwonot when testing hypothesis \hyppath.

\subsubsection{Protocol adjustments: participants \LV}
\label{sec:protocole-adjustements-lv}
For participants \LV, pilot sessions revealed that 6 testing trials were overly demanding for this group: several participants suddenly experienced acute visual fatigue, making it impossible to continue and leading to early interruptions. As a consequence, this protocol design risked unbalancing the data, with many completing only a single condition. To address this, we reduced and reorganized the trials and blocks (changes in Fig~\ref{fig:protocole-overview} (4) and (5)):
\begin{itemize}
    \item Participants first completed two trials per condition, each followed by a NASA-TLX questionnaire, to ensure that subjective workload was captured before cumulative fatigue could bias responses.
    \item The order of layouts was randomized but kept identical across \condonenot and \condtwonot to support fair comparisons.
    \item To maximize usable data, participants continued with the remaining four trials per condition in an alternating sequence (one \condonenot, one \condtwonot, and so on). Sessions ended as soon as all trials were completed, the participant indicated they were too tired to continue, or the 30-minute time limit was reached.
    \item Finally, participants reported their overall preference between \condonenot and \condtwonot, or both equally. Because cumulative fatigue strongly influenced judgments, we also administered a comparative NASA-TLX at the end of the session to balance condition evaluations.
\end{itemize}

Having outlined the overall procedure, the following subsections expand on its key components: the creation of newspaper page material, the conditions established based on CPS, the population, and the data collected for subsequent analysis.

\subsection{Newspaper pages material}
\label{sec:newspaperpages}

\subsubsection{Original edition}

Each newspaper page combined two components: (1) a \emph{layout file}, defining the geometric arrangement of articles on the page, and (2) a \emph{content file}, providing the textual material for each article (headline, body text, and target font sizes). Because our study focuses on headlines, body text was generated only to fill space within article blocks and was not analyzed.  

\begin{figure}[htbp]
    \centering
    \includegraphics[width=\widthForOneImage]{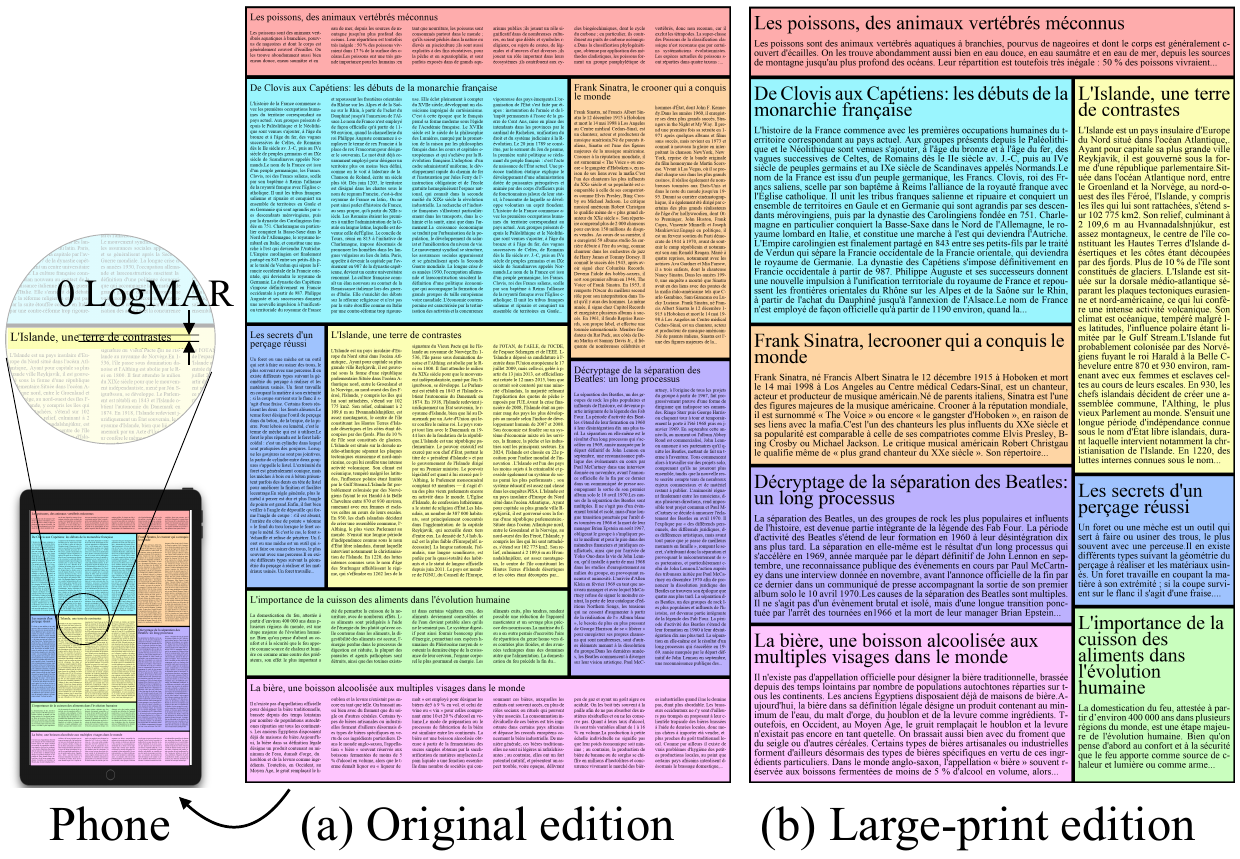}
    \caption{\label{fig:newspaper-pages}
    Examples of newspaper pages: (a) original edition used in condition \condonenot, and (b) corresponding large-print edition used in condition \condtwonot, automatically generated using the re-layouting method from Gallardo et al.~\cite{gallardo} with a $\times 2$ magnification.  
    Colors indicate matching articles across the two versions (i.e., occupying the same area on the page). The textual content differs between (a) and (b) to avoid repetition effects, but was balanced in headline length. On the left-hand side, we also illustrate how the original edition would look in a phone, which is the display used to define the headline size in LogMAR.}
    \Description{The figure shows two versions of the same newspaper page. Panel (a) presents the original edition, while panel (b) shows the corresponding large-print edition automatically generated using a re-layouting algorithm with a twofold magnification. Matching articles across the two versions are highlighted with colors to indicate their corresponding positions, though the textual content differs to avoid repetition effects; headline lengths were kept balanced. Additionally, panel (a) includes a depiction of the original edition as it would appear on a phone, illustrating the device and the headline size in LogMAR for our study.}
\end{figure}

In total, we generated 18 newspaper pages, each containing between 8 and 11 articles (see \figref{fig:newspaper-pages}(a)). Below, we detail the construction of layouts, headline sizes, and content.

\paragraph{Font size.}  
Headline size was fixed to correspond to a visual angle of $0$ logMAR when displayed on the phone at a viewing distance of 40\,cm (\figref{fig:newspaper-pages}(a); see \secref{sec:condition-definition}). This $0$ logMAR size was intentionally chosen to challenge normally-sighted participants, as it requires zooming for comfortable reading~\cite{calabrese_baseline_2016}. Body text size was set heuristically to half the headline size. 

\paragraph{Layouts.}  
Layouts were created manually from annotated front pages of the \NYT. We produced nine layout files ($\layoutset=\{l_i\}_{i=1,..,9}$), each specifying the number of articles, their position, and their dimensions (width and height in mm). Based on these dimensions and the font size, we extrapolated the required article length in characters and lines. 

More precisely, concerning headlines, to match the New York Times design, we first analyzed headline lengths relative to article width (number of columns). This showed that headlines in narrower articles (spanning fewer than two columns) typically used up to three lines, while those in wider articles (three or more columns) almost always fit on a single line. We reproduced this distribution when generating our own headlines (see example in \figref{fig:newspaper-pages}(a)). 

\paragraph{Content.}  
Given article length specifications, we created three distinct content files for each layout file, yielding a total of 27 sets of article content $\contentset=\{c_{ij}\}_{i=1...9, j=1,2,3}$. This ensured that participants were not exposed to the same material across conditions (\condzeronot, \condonenot, \condtwonot), avoiding memory bias. Content files were created in R~\cite{R} by selecting the article's text from the web. For each article, a matching headline was then generated automatically with MistralAI~\cite{mistralai}. \addedimx{We utilized this LLM-based approach to generate controlled media stimuli with uniform linguistic difficulty. This ensured that the behavioral integrity of the reading paths was measured against standardized content, eliminating potential engagement biases or emotional confounding that could arise from prior knowledge of real-world news events.}

As a starter, a list of 364 key words (e.g., dog, Paris, etc.) was manually written to serve as article topics and sorted within 30 general themes (e.g., animals, cities, etc.). For each keyword, the R~\cite{R} \emph{getwiki} library was used to search the web and extract the first paragraph of the top 20 Wikipedia and Vikidia pages (in French) matching the search result. This resulted in a list of 11,565 articles. Using a custom-based routine in R, this list was searched to assign the optimal candidate articles to each layout file, based on the required character length, checking that the same theme was only chosen once within a single page to avoid redundant topics. 

Last, MistralAI was prompted from R to generate a list of 10 titles per article automatically. The prompt sent to MistralAI was as follows: "Create 10 possible headlines in the style of a journal article, in French, with approximately \emph{n} characters (space included), about the following article \emph{t}", \emph{n} being the number of characters given by the layout files and \emph{t} being the text of the article. Each output was manually inspected to select the optimal headline. 

\subsubsection{Large-print edition}

For each original edition, we generated a large-print counterpart by applying the automatic re-layouting method from Gallardo et al.~\cite{gallardo}. \addedimx{This approach addresses the limitations of simple font scaling (which can cause text overflows and break entry points because original layouts were not designed for increased sizes~\cite{gallardo}).} This method \added{takes an original edition and a font size magnification factor to apply, and} uses an evolutionary algorithm to reorganize layouts in a way that optimizes the number of lines of each headline and the aesthetic quality of the layout e.g., alignment of articles, visual balance, etc. \addedimx{It is crucial to note that the goal of this adaptation is structural integrity over the preservation of absolute coordinates. While the articles are re-positioned to accommodate the larger font size, the 2D layout topology is maintained.}

We feed this method with a layout $l_i$ and one content set ($c_{i,1}$), together with a magnification factor of $\times 2$, which comfortably covers the critical print size spectrum across both participant groups (see \secref{sec:condition-definition}). \deleted{and also allows to fully display the headline, which is important for the comparative analysis between conditions.}  Only $c_{i,1}$ was used for generating these layouts (since $c_{i,2}$ and $c_{i,3}$ are equivalent in terms of length). The output is a new large-print layout $\largeprintset=\{l'_i\}_{i=1,...,9}$ (\figref{fig:newspaper-pages}(b)). The large-print edition was obtained by combining the large-print layout with one of the three equivalent content files for that page. Note that in the large-print edition, a larger font size inevitably prevents the full body text of each article from being displayed in the overview. This is an inherent feature of these large-print digital editions, which, like existing kiosk-style news applications, are designed primarily to support scanning and discovery of articles. Readers would use the overview to identify entry points (e.g., headlines), and then would access the full content in a dedicated single-page reading mode.

\subsection{Conditions setup based on CPS}
\label{sec:condition-definition}

To recap, our goal is to design an experiment where condition \condonenot forces participants to use gesture-based magnification, as the original edition’s headlines are deliberately too small to read comfortably, while condition \condtwonot allows participants to read the large-print edition without any magnification. Importantly, for each participant, both conditions are presented on the same device and at the same head-to-screen distance. The procedure to achieve this setup is detailed in the following steps.

\subsubsection{Estimation of the CPS}

To assess participants’ CPS, we used a standardized reading test based on the MNREAD chart~\cite{legge1992psychophysics,mansfield1993}. The MNREAD is a continuous-text reading acuity chart consisting of short sentences presented at progressively smaller print sizes. Reading speed is plotted as a function of print size, producing the characteristic MNREAD curve: a plateau of constant reading speed at larger sizes, followed by a sharp decline once print falls below a critical threshold. The CPS is defined as the smallest print size that still supports maximum reading speed, i.e., the lower bound of the plateau~\cite{legge2006}. \deleted{CPS is expressed in logMAR units (logarithm of the Minimum Angle of Resolution), which
quantify text size in terms of visual angle rather than physical dimensions. Using visual angle is crucial, as it jointly
depends on both print size in millimeters and viewing distance—the two factors that ultimately determine legibility.}

\subsubsection{Headline size in LogMAR for both conditions}

\begin{figure}[htbp]
    \centering
    \includegraphics[width=\widthForOneImage]{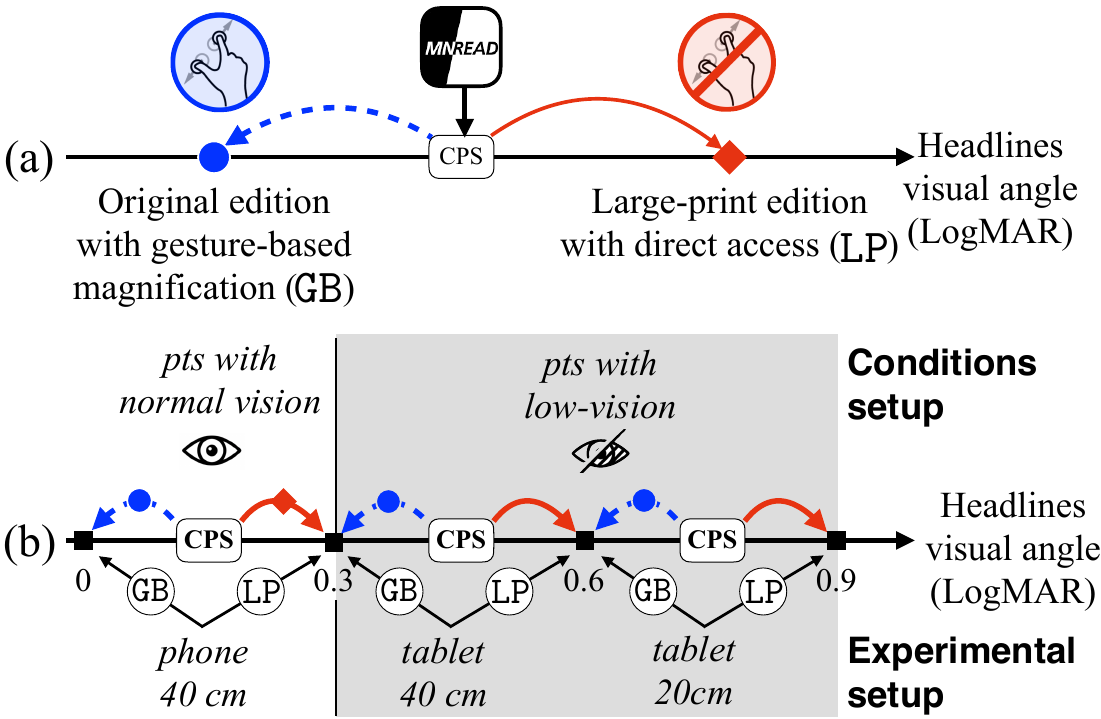}
    \caption{\label{fig:conditions-setup}
    Experimental condition setup based on CPS.  
    (a) General principle: headline visual angles in both conditions are selected around the participant’s CPS value obtained from an MNREAD test.  
    (b) Device and viewing distance configuration depending on the scenario: for participants \NV ($0 \leq CPS \leq 0.3$), the experiment was conducted on a phone at a fixed distance of 40\,cm; for participants \LV ($0.3 \leq CPS \leq 0.9$), the experiment was conducted on a tablet, with the viewing distance adjusted according to the CPS value.}
    \Description{The figure shows how headline sizes in both \condonenot and \condtwonot were adapted to each participant’s Critical Print Size (CPS). Panel (a) illustrates that \condonenot and \condtwonot conditions are set relative to CPS in terms of headline visual angle. On the x-axis representing headline visual angle, it shows that the two conditions are positioned on either side of the participant’s CPS. Panel (b) uses the same representation but instead of illustrating the principle, it shows the actual device and viewing distance configurations chosen for the experiment according to different CPS ranges. Participants \NV used a smartphone at 40\,cm, while participants \LV used a tablet with distance adjusted to their CPS. This ensures that \condonenot headlines require magnification, while LP headlines are directly legible.}
\end{figure}

The visual angle of headlines depends on two factors: the device’s screen size and the viewing distance. Given a participant’s CPS, the goal is therefore to select the appropriate device and distance so that the headline size of the original edition in \condonenot is smaller than the CPS, while the headline size of the large-print edition in \condtwonot is larger than the CPS. When multiple configurations are possible, we systematically choose the one that yields headline sizes closest to the CPS. This principle is illustrated in \figref{fig:conditions-setup}(a).

A custom-designed app, developed using React Native framework, was deployed for the application of this protocol.
Regarding this, we use two different Android devices:
\begin{itemize}
    \item \textbf{Phone:} Google Pixel~7, featuring a 7.3-inch screen with a maximum resolution of $1080 \times 2400$ pixels.  
    \item \textbf{Tablet:} Samsung Galaxy Tab S9 FE+, featuring a 12.4-inch screen with a maximum resolution of $1600 \times 2560$ pixels.  
\end{itemize}

These specifications, combined with adjustments of the viewing distance, allowed us to cover participants with CPS values ranging from 0 to 0.9. The actual configuration depended on each participant’s CPS, as illustrated in \figref{fig:conditions-setup}(b). For participants \NV, we used the phone at a fixed viewing distance of 40\,cm. For participants \LV, we used the tablet, with the viewing distance adjusted according to their CPS.

\subsection{Population}

We recruited 24 participants: 19 participants with normal or corrected-to-normal vision and 5 participants \LV. The sample size meets \addedcr{SIGCHI} standards for behavioral studies~\cite{caine-2016}. \addedcr{The number of participants \LV ($n=5$) is considered acceptable when conducting research with users with disabilities~\cite{lazar-hci-2017}}

 Participants \NV were mainly recruited through an open survey, while participants \LV were recruited from the \addedcr{CHU Pasteur ophthalmology department}. All participants took part voluntarily. Demographic details for both groups are provided in Table~\ref{tab:population}.

\begin{table}[htbp]
    \centering

\begin{tabularx}{\columnwidth}{rlllll}
\toprule
ID & \makecell[b]{Age\\ group} & Gender & \makecell[b]{CPS \\ (LogMAR)} & \makecell[b]{Reading \\ frequency}  & Diagnosis \\
\midrule
1 & 20-35 & Male & 0.15 & Daily & NA \\
2 & 20-35 & Female & 0.1 & Rarely & NA \\
3 & 20-35 & Female & 0.1 & Never & NA \\
4 & 20-35 & Male & 0.15 & Daily & NA \\
5 & 20-35 & Male & 0.1 & Daily & NA \\
7 & 20-35 & Male & 0.15 & Daily & NA \\
8 & 20-35 & Female & 0.2 & Daily & NA \\
9 & 36-65 & Male & 0.15 & Daily & NA \\
10 & 20-35 & Male & 0.1 & Weekly & NA \\
11 & 20-35 & Male & 0.2 & Rarely & NA \\
12 & 20-35 & Male & 0.1 & Rarely & NA \\
13 & 36-65 & Female & 0.1 & Daily & NA \\
14 & 36-65 & Male & 0.1 & Daily & NA \\
15 & 20-35 & Female & 0.1 & Weekly & NA \\
16 & 20-35 & Male & 0.1 & Weekly & NA \\
17 & 20-35 & Male & 0.08 & Weekly & NA \\
18 & 66-80 & Male & 0.1 & Daily & NA \\
19 & 36-65 & Female & 0.2 & Daily & NA \\
20 & 66+ & Female & 0.15  & Daily &  NA\\
\rowcolor{gray!25}21 & 80+ & Male & 0.8 & Daily & Glaucoma \\
\rowcolor{gray!25}22 & 80+ & Male & 0.5 & Daily & Glaucoma \\
\rowcolor{gray!25}23 & 80+ & Male & 0.45 & Weekly & Glaucoma \\
\rowcolor{gray!25}24 & 66-80 & Female & 0.9 & Never & AMD 
\\
\rowcolor{gray!25}25 & 66-80 & Female & 0.9 &Daily &  AMD \\
\bottomrule
\end{tabularx}
    \caption{\label{tab:population}
    Participant demographics, CPS, and newspaper reading frequency. The last column lists diagnoses for participants \LV, with corresponding rows shaded in light gray. Skipped ID corresponds to a dropped participant because of unusable data.}    
\end{table}

\subsection{Data collected and measures}

For each participant, we recorded the following data during the experiment \addedimx{to provide a multidimensional assessment of the quality of experience}:
\begin{itemize}
    \item \textbf{Pre-experiment questionnaire responses:} Age group, gender (optional), and news reading habits.
    \item \textbf{Critical Print Size:} CPS in logMAR units, obtained from the MNREAD reading test.
    \item \textbf{Task timing:} Total duration of each task in each trial (reading headlines, finding target articles), measured in seconds by the application.
    \item \textbf{Task success:} For each trial, the percentage of read articles in task 1, and also a boolean indicator for whether the participant correctly found the target article in task 2. Both of these for each trial. 
    \item \textbf{Voice recordings:} Audio of participants reading aloud headlines, enabling reconstruction of reading order.
    \item \textbf{Screen interactions:} For \condonenot trials, detailed logs of gestures, including zoom level, pan position, timestamps, and sequence, allowing reconstruction of reading paths.
    \item \textbf{Subjective workload:} Responses to the NASA-TLX questionnaire were proposed after each condition.
    \item \textbf{Overall condition preference:} Participant choice between \condonenot and \condtwonot, or both equally preferred.
\end{itemize}
For participants \LV, time limits and adapted trial sequences were applied as described in \secref{subsection:overview}, and all data were recorded consistently across conditions to enable fair comparisons.

Note that, except for voice recordings, no personal data was collected. Voice recordings were stored temporarily on an encrypted laptop and processed locally using a semi-automatic speech-to-text service to extract reading order and timestamps. Once transcription was completed, the recordings were permanently deleted.

\section{Results} \label{sec:results}

Our analysis tested hypotheses \hypperformance-\hyppreference.  We first introduce the statistical model, then report results for Task 1 and Task 2, and finally analyze subjective workload and user preference using the NASA-TLX questionnaire together with overall preference.

\subsection{Statistical model}

We analyzed the data using a linear mixed-effects model with Participant ID as a random effect. The fixed effects are detailed in Table~\ref{tab:lmm}. We ran an ANOVA to test for significance using the F-statistic, setting the p-value threshold at $p < 0.05$. All reported values are model estimates (i.e., estimated marginal means) rather than raw data. 

The following sections report exact p-values and effect sizes for significant main effects and interactions. Complete model statistics—including F-values, degrees of freedom, and detailed pairwise comparisons—are provided in the supplementary tables in the Appendix. \addedcr{We do not report main effects and interactions that were not significant}.

\begin{table}[htbp]
\centering
\setlength{\tabcolsep}{3pt}
\begin{tabularx}{\columnwidth}{lXXX}
\toprule
\textbf{\#} & \textbf{Factor (\#Levels)} & \textbf{Levels} & \textbf{Interactions} \\
\midrule
1 & Condition (2) & \condonenot, \condtwonot & With \#2, \#3, \#4, \#5, \#8 \\
2 & NewspaperID (6) & 1,...,6 & With \#1, \#3 \\
3 & \CVAS (2) & Small Screen, Low Vision & With \#1, \#2 \\
\rowcolor{gray!20} 
4 & Reading frequency (4) & Daily, Weekly, Rarely, Never & With \#1 \\
\rowcolor{gray!20}
5 & Target headline zone (3) & Top, Middle, Bottom & With \#1 \\
\rowcolor{gray!20}
6 & Trial index (6) & 1,$\ldots$, 6 & Main effect only \\
\rowcolor{gray!20}
7 & Content version (3) & 1, 2, 3 & Main effect only \\
\rowcolor{gray!20}
8 & Condition order (2) & \condonenot{} first, \condtwonot{} first & With \#1 \\
\bottomrule
\end{tabularx}
\caption{\label{tab:lmm}
Fixed effects of the Linear Mixed-Effects Model (LMM) used in this study. Factor of interests are shown with a white background; secondary or control factors in gray.}
\Description{This table outlines the eight fixed effects used in the study's Linear Mixed-Effects Model (LMM). For each effect, the table lists the factor's name, its number of levels, a description of those levels, and its interactions with other factors. The factors are grouped into two categories.
First, the main Factors of interest are Condition (2 levels; \condonenot and \condtwonot), NewspaperID (6 levels; numbered 1 through 6), and \CVAS (2 levels, "Small Screen" and "Low Vision."). 
Then we describe the control Factors, namely: 
Reading frequency (4 levels; Daily, Weekly, Rarely, Never), 
Target headline zone (3 levels; Top, Middle, Bottom),
Trial index (6 levels; Numbered 1 through 6; main effect only),
Content version (3 levels; 1, 2, 3; main effect only), and
Condition order (2 levels; Which of the two main conditions was presented first). The "Interactions" column specifies which factors were modeled to influence each other. For example, the main Condition factor was modeled with interactions for nearly all other factors.}
\end{table}

\subsection{Task 1 analysis}
\begin{figure}[htpb]
    \centering
    \includegraphics[width=\widthForOneImage]{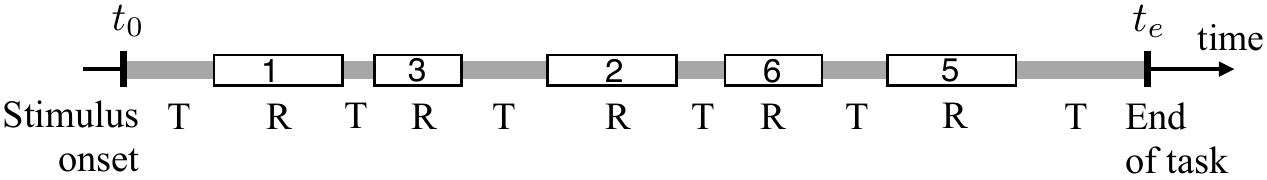}
    \caption{\label{fig:taskone-timeline}
Task 1 timeline.  
After stimulus onset (time $t_0$, when the newspaper page is displayed), participants alternate between transition periods for locating the next headline (T) and reading periods (R, with white boxes indicating the headline index). The task ends at time $t_e$, when the participant indicates they have read all headlines—even if some were missed, as illustrated in this example where headline 4 was skipped. 
    }
    \Description{The figure depicts the timeline of Task 1, showing the sequence of actions a participant performs while reading a newspaper page. Each reading period (R) corresponds to the time spent on a specific headline, with its index shown in a white box. Transition periods (T) represent the time spent searching for the next headline. The timeline begins at stimulus onset ($t_0$) and ends when the participant signals completion ($t_e$). This visualization highlights the alternation between reading and transition and shows that participants may sometimes skip headlines during the task.}
\end{figure}

We first transcribed participants’ audio recordings using WhisperX~\cite{whisperX}, reconstructing a task timeline for each participant. Each timeline was segmented into reading (R) and transition (T) periods (see \figref{fig:taskone-timeline}). From this timeline, we derived four metrics to test hypotheses \hypperformance and \hyppath:
\begin{enumerate}
    \item Success ratio: The proportion of correctly read articles out of the total (e.g., 5/6=$83.3\%$ in \figref{fig:taskone-timeline}). Note that a headline was considered correctly read if its transcription was phonetically equivalent to the source text, which we verified using the \emph{phonetic-fr} Python library.
    \item Completion time: The total task duration, from stimulus onset to completion ($t_e - t_0$ in \figref{fig:taskone-timeline}). 
    \item Reading path: The sequence in which articles were read (e.g., 1-3-2-6-5 in \figref{fig:taskone-timeline}).
    \item Reading and transition time: The durations of individual reading and transition periods.
\end{enumerate}

\subsubsection{Success ratio}
Our analysis revealed significant effects for three factors:
\begin{itemize}
\item {NewspaperID:} Although significant differences existed between newspapers \added{(\statreport{5}{215.94}{5.72}{<.001}{0.12})}, performance was consistently high, with the lowest-performing newspaper still achieving a $93.4\%$ success ratio.
\item {CVAS:} Participants \NV achieved a higher success ratio ($100\%$) than participants \LV ($93.4\%$), consistent with the reduced vision in the latter group \added{(\statreport{1}{23.71}{20.92}{<.001}{0.55})}.
\item {Content version:} A significant difference was found between Content \#1 ($96.2\%$) and \#3 ($98.6\%$), but both yielded high success ratios \added{(\statreport{2}{223.45}{3.72}{=.025}{0.03})}.
\item \added{A significant interaction between CVAS and NewspaperID (\statreport{5}{216.1}{3.83}{=.002}{0.08}) indicated that the performance gap due to vision loss was layout-dependent, something not present in Layouts 16 and 17.}
\end{itemize}

Most importantly, we found no significant effect of Condition, indicating that neither \condonenot nor \condtwonot affected the participants' ability to read headlines correctly.

\subsubsection{Completion time}
Our analysis revealed significant effects for all our factors of interest:
\begin{itemize}
    \item {Condition:} The main result is that, \addedimx{as anticipated due to the reduction in navigational overhead}, \condtwonot had $37\%$ lower completion time than \condonenot \comparison{$71.7$\seconds}{$45.2$\seconds} \added{(\statreport{1}{209.43}{116.5}{<.001}{0.42})}. An interaction with \CVAS \added{(\statreport{1}{209.18}{146.3}{<.001}{0.41})} showed that this improvement was larger for  participants \LV, with a $42\%$ reduction \comparison{$114.7$\seconds}{$66.3$\seconds} compared to a reduction of $16\%$ for participants \NV \comparison{$28.6$\seconds}{$24.1$\seconds}). This finding supports Hypothesis~\hypperformance.

    \item {\CVAS}: Individuals \NV (average of $26.3$\seconds) had $70\%$ lower completion times than readers \LV (average of $90.5$\seconds), consistent with the reduced vision in the latter group \added{(\statreport{1}{18.61}{50.19}{<.001}{0.73})}. 

    \item {NewspaperID:} Layout \#17, featuring a feed-like format (i.e., articles displayed as a stack of rectangles, as commonly seen in social media feeds) had significantly shorter completion time (avg. $48.9$\seconds) than the others (avg. $60.3$\seconds) \added{(\statreport{5}{209.89}{9.57}{<.001}{0.19})}. While effective, we consider this a special case with limited generalization (further discussed in \secref{subsection:preferences}).

    \item {Content version:} A significant difference \added{(\statreport{2}{213.5}{3.21}{=.04}{0.03})} was found between version \#1 (avg. $55.9$\seconds) and \#3 (avg. $61$\seconds), likely due to subtle variations in content complexity that influenced processing time or even induce more mistakes.

\end{itemize}

\subsubsection{Reading path}

To analyze reading paths, we first identified a reference reading strategy in \condzeronot trials. Visual inspection initially suggested a left-to-right, top-to-bottom reading pattern in \condzeronot trials. We then quantified the similarity between each path and this candidate pattern using a similarity measure\footnote{Given two paths $p_1$ and $p_2$, we defined their similarity as $Sim(p_1,p_2) = 100*(1.0 - levenshtein(p_1,p_2)/max(length(p_1), length(p_2)))
$, using the Levenshtein distance~\cite{levenshtein2007}. The similarity  $Sim(p_1,p_2)$ ranges from 0, indicating completely different paths, to 100, indicating $p_1 = p_2$.} and an ANOVA confirmed no significant deviation from it. Based on this validation, we define the {\itshape reference path} as the theoretical left-to-right, top-to-bottom reading order implied by the page layout, independent of any experimental data.

Using the reference paths, we computed the similarity of \condonenot and \condtwonot reading paths against reference paths induced over the corresponding layouts.  \addedimx{To enable sequence comparison, we indexed the articles in each condition (\condonenot and \condtwonot) according to the reference strategy (e.g., Article \#1 is the top-leftmost). Consequently, a path adhering to the reference strategy results in the canonical sequence [1,2,...,n], while any deviation will be captured by the similarity measure.}

\added[comment="3. Reading Path"]{It is important to note that we do not treat deviation from this reference path as a performance failure or a "negative" outcome. Rather, we interpret such deviations as behavioral evidence of a strategic shift in reading, quantifying the extent to which each condition disrupts the structural perception of the 2D layout.} 

\begin{figure}[htpb]
    \centering
    \begin{tabular}{ccc}
    \includegraphics[width=\widthForThreeImages]{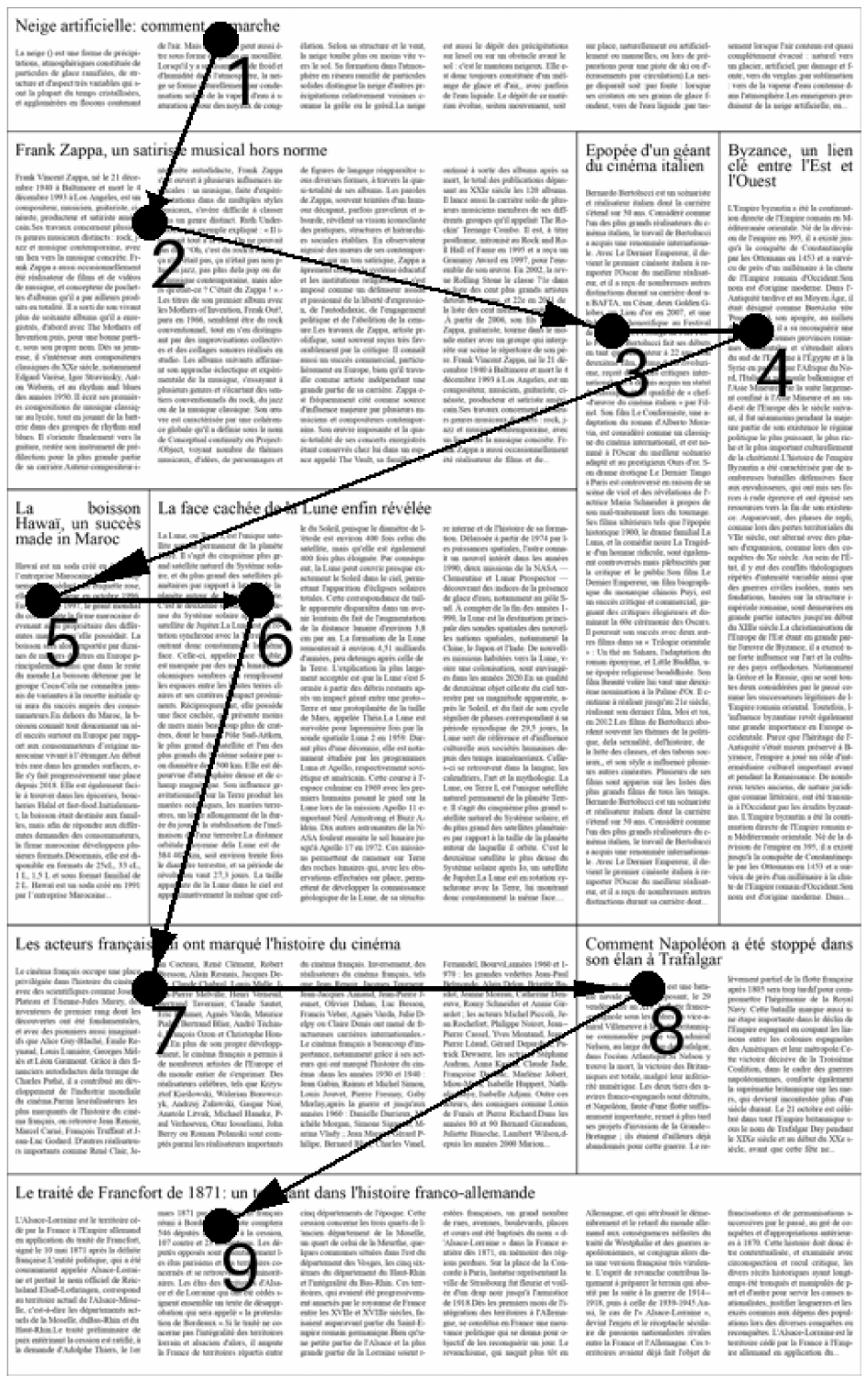}&  \includegraphics[width=\widthForThreeImages]{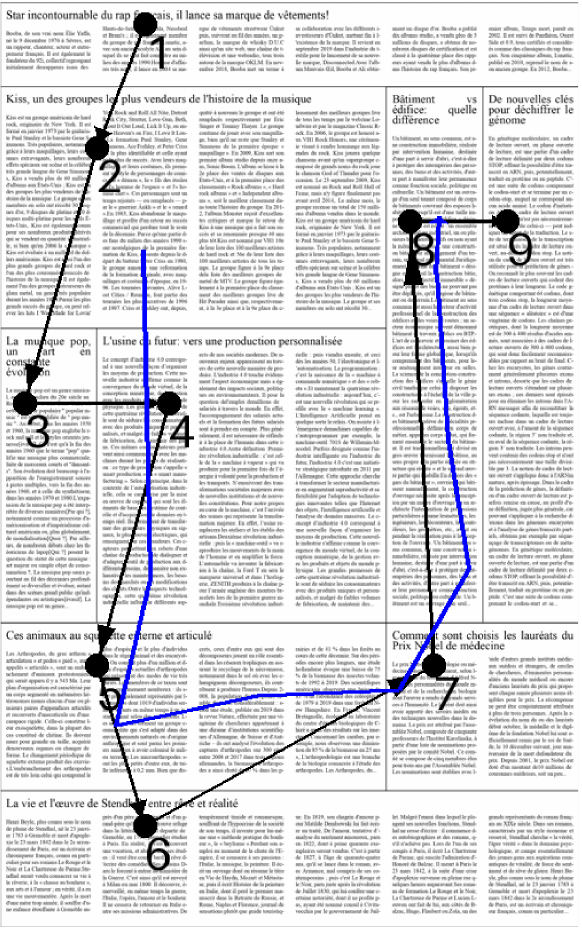}& \includegraphics[width=\widthForThreeImages]{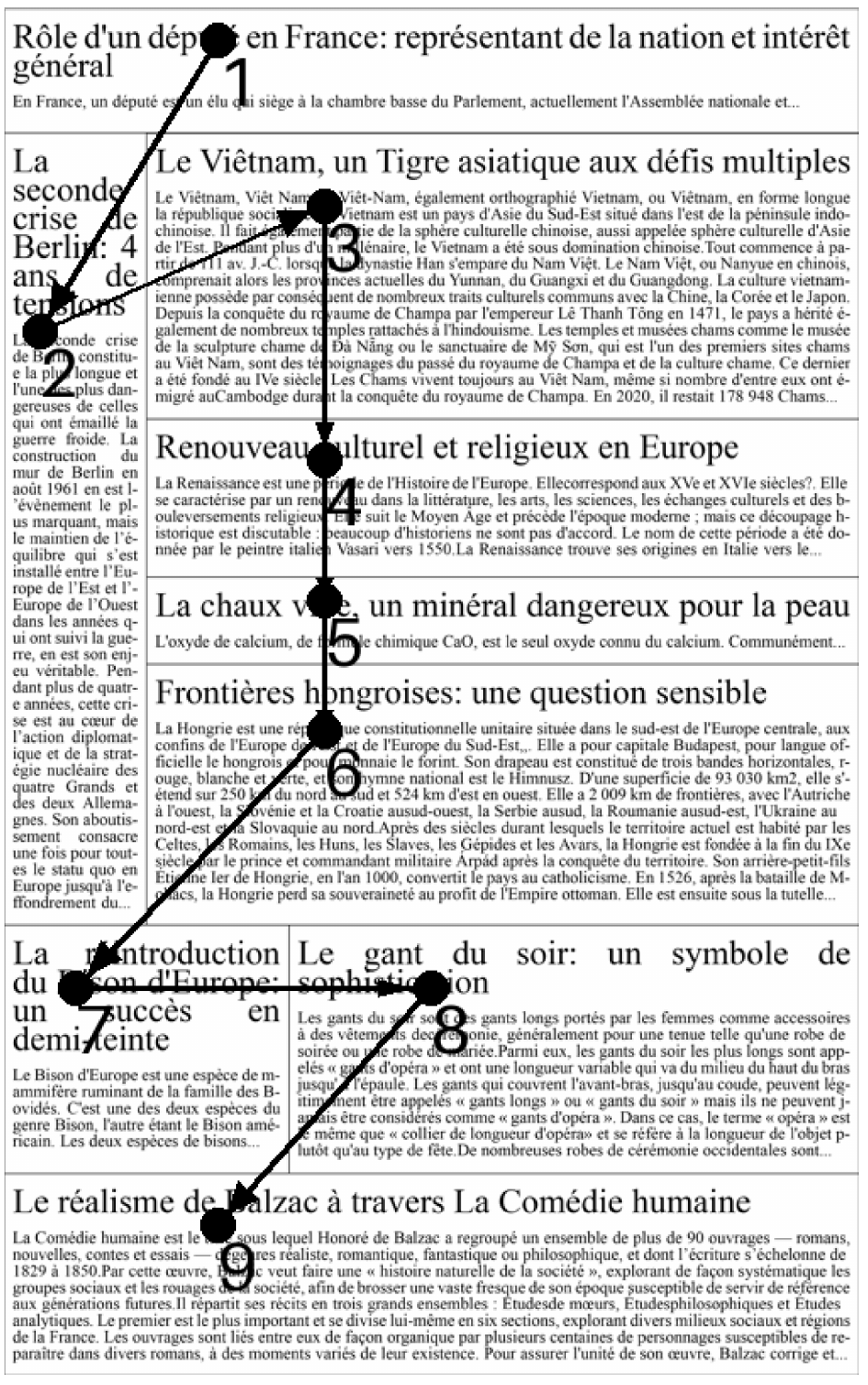}\\[-1mm]
    (a) \condzeronot & (b) \condonenot & (c) \condtwonot\\[-1mm]
    \end{tabular}
    \caption{
    \label{fig:example-reading-path}
    Examples of reading path corresponding to the same participant and layout in all conditions. 
    (a) \condzeronot path, reassembling always the reference path (top-to-bottom, left-to-right). 
    (b) \condonenot path, with the screen interactions i.e., panning movements (in blue). Zoom level is not shown to simplify the figure and emphasize gesture patterns.
    (c) \condtwonot path.
    }
    \Description{Figure~\ref{fig:example-paths} illustrates reading paths from the same participant and layout across the three conditions. In \condzeronot (a) and \condtwonot (c), the path follows a strategy close to the reference path (left-to-right, top-to-bottom order). In contrast, the path in \condonenot (b) shows deviations from this reference pattern; instead, it shows similarities to the screen interaction pattern given by the panning gestures.}
\ \\[2mm]
   \centering
    \begin{tabular}{cccc}
    \includegraphics[width=\widthForFourImages]{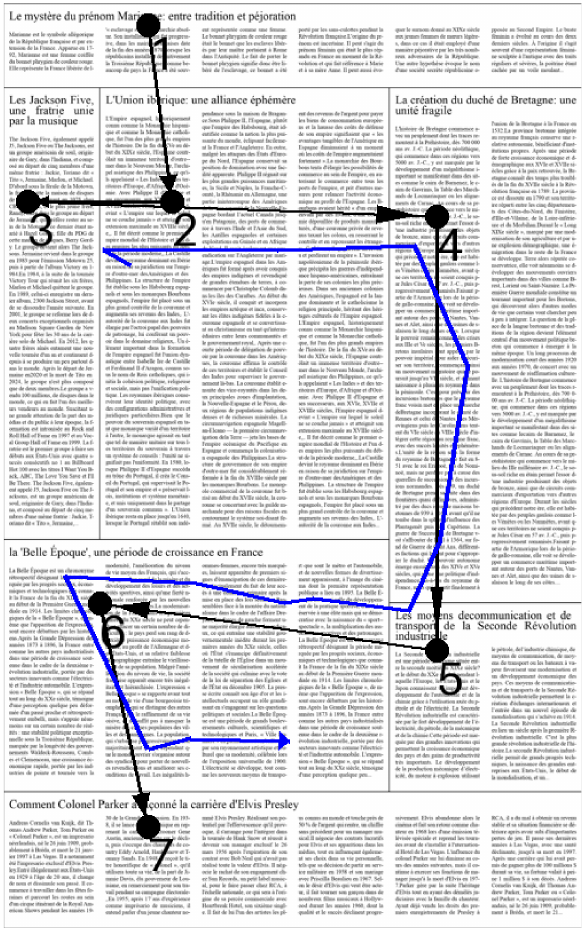} & 
    \includegraphics[width=\widthForFourImages]{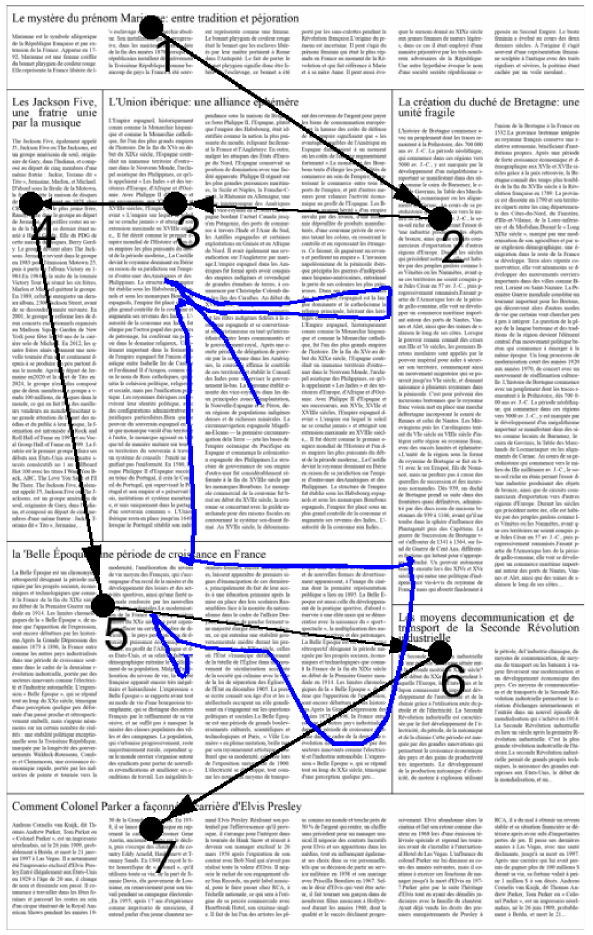} &
    \includegraphics[width=\widthForFourImages]{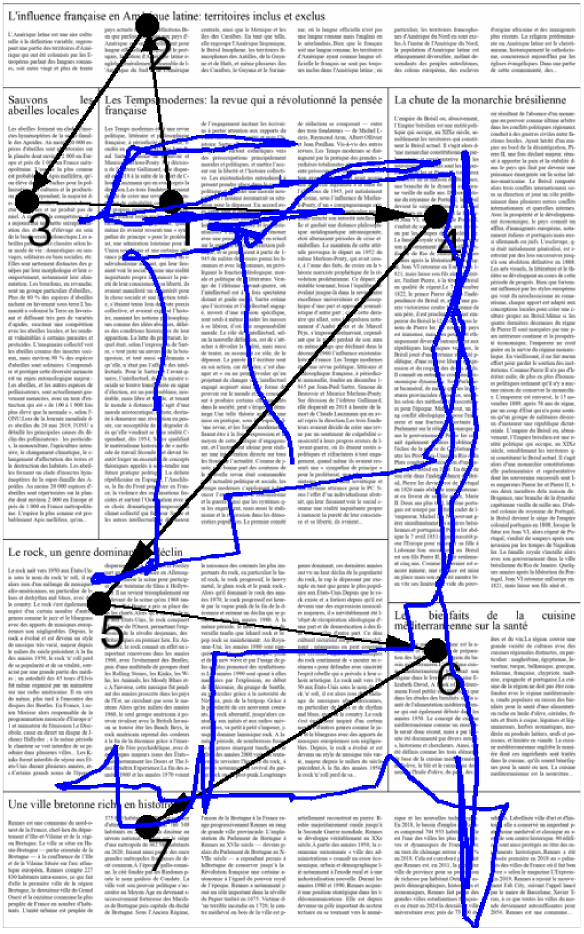} & 
    \includegraphics[width=\widthForFourImages]{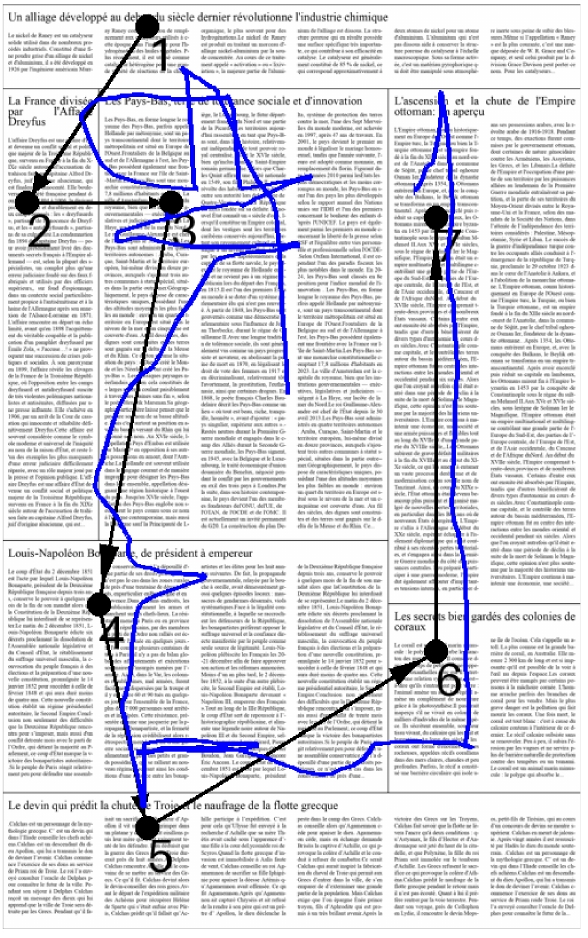} \\[-1mm]
    (a) & (b) & (c) & (d)
    \end{tabular}
    \caption{\label{fig:example-screen-gestures}
    Four examples of \condonenot reading paths (black) with screen interactions (panning movements in blue) for the same layout (\#17). In \condonenot, reading follows panning gestures, which do not necessarily align with the reference path. (a,b) show two different participants \NV. (c,d) show two participants \LV, exhibiting similar behavior but with considerably more gestures, indicating that the task is more demanding for this group.
    }
    \Description{Figure~\ref{fig:example-screen-gestures} shows four examples of how users navigated a digital newspaper page using gesture-based magnification. Each of the four panels displays the same newspaper layout.On each panel, two types of paths are shown:

    A black line with numbered points and arrows indicates the reading path, connecting headlines in sequential order.

    A thick blue line shows the user's actual panning movements (screen interactions) as they explored the page.

Panels (a, b) show the paths for two participants with normal vision. Their blue panning paths consist of relatively simple horizontal and vertical gestures that follow the content.

Panels (c, d) show the paths for two participants with low vision. Their blue panning paths are significantly denser and more complex, with many more movements. This visually demonstrates that the task was considerably more effortful and demanding for them.}
\end{figure}

Our analysis revealed three significant factors:

\begin{itemize}
    \item {Condition:} The main effect observed was that reading paths in \condtwonot were significantly more similar to the reference than in \condonenot \comparison{$69.3\%$}{$87.5\%$} \added{(\statreport{1}{210.75}{11.98}{<.001}{0.14})}. This effect was present for both participants \NV \comparison{$72.3\%$}{$88.9\%$} and for participants \LV \comparison{$66.2\%$}{$86.1\%$}. The lower similarity in \condonenot \replacedimx{reveals a measurable behavioral cost: while manual interaction provides control, the navigational overhead required to manage successive zoomed-in fragments disrupts the reader's ability to maintain their natural exploration strategy.}{ is likely due to the influence of the local field of view: while manipulable interaction offers control, it restricts global awareness, leading participants to prioritize articles visible within the current view rather than following the global document structure.}
    This effect is visible in \figref{fig:example-screen-gestures}, where participants \deleted{\NV} exhibited reading paths guided by their screen interactions (i.e., panning movements), and in particular, participants \LV required substantially more gestures, reflecting the greater effort. This observation supports hypothesis~\hyppath.

    \item {NewspaperID:} Layout influenced the reading path \added{(\statreport{5}{212.51}{6.15}{<.001}{0.13})}. While \condtwonot generally produced paths more similar to the reference, \added{a significant interaction showed that} \added{(\statreport{5}{209.19}{8.93}{<.001}{0.18})} layout \#19 (\figref{fig:newspaper-pages}) was a notable exception, where \condonenot scored slightly higher \comparison{$72.2\%$}{$68.6\%$}. This may be due to its unique vertical organization. However, the main effect of \condtwonot superiority held across all other layouts.

    \item {Reading frequency:} We found an interaction between condition and reading habits \added{(\statreport{3}{212.49}{5.94}{<.001}{0.08})}. In \condonenot, frequent readers (daily or weekly) adhered more closely to the reference path than infrequent readers (who averaged only $51.4\%$ similarity). In contrast, \condtwonot paths were highly similar to the reference (over $83\%$) across all reading frequencies, demonstrating their robustness.
\end{itemize}

\subsubsection{\replaced{Participant-level performance}{Reading path similarity and completion time}}

\begin{figure}[htpb]
    \centering
    \begin{tabular}{ccc}
    & {\bf Absolute} & {\bf Relative}\\[1mm]
    \raisebox{25pt}[0pt][0pt]{\rotatebox{90}{{\bf \nv}}} &
    \includegraphics[width=0.45\widthForOneImage]{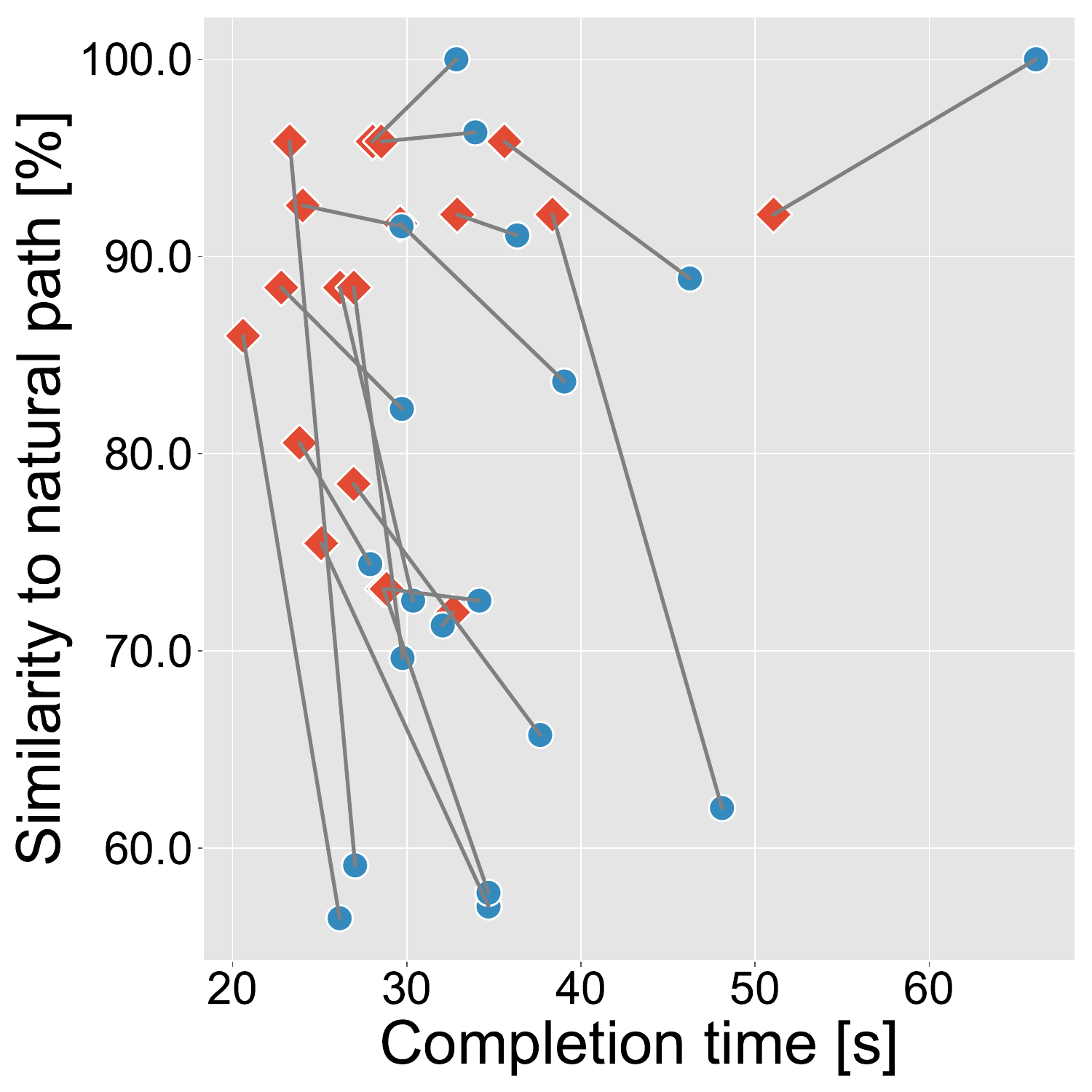} & 
    \includegraphics[width=0.45\widthForOneImage]{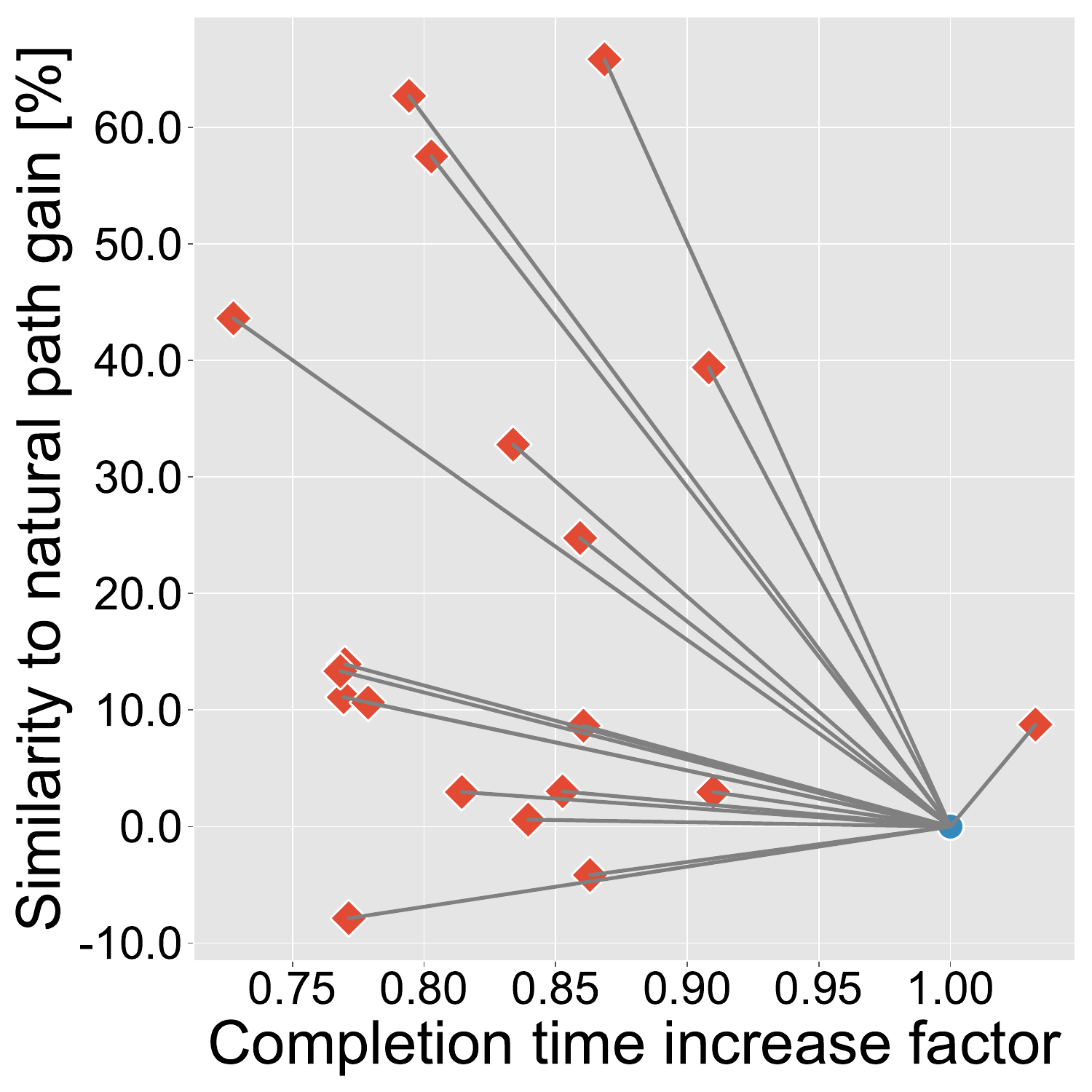} \\[-1mm]
    & (a) & (b) \\[1mm]
    \raisebox{25pt}[0pt][0pt]{\rotatebox{90}{{\bf \lv}}} &
    \includegraphics[width=0.45\widthForOneImage]{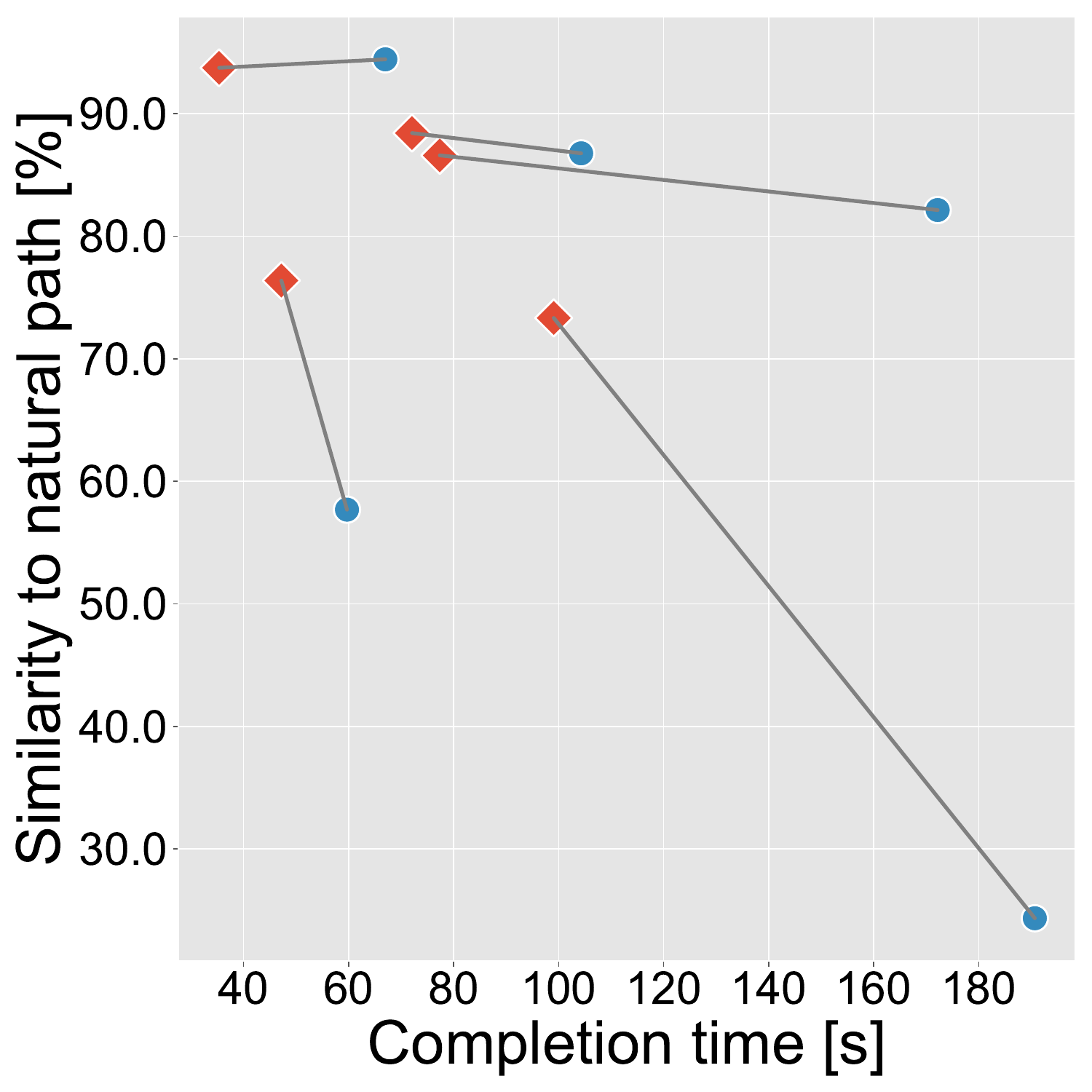} & 
    \includegraphics[width=0.45\widthForOneImage]{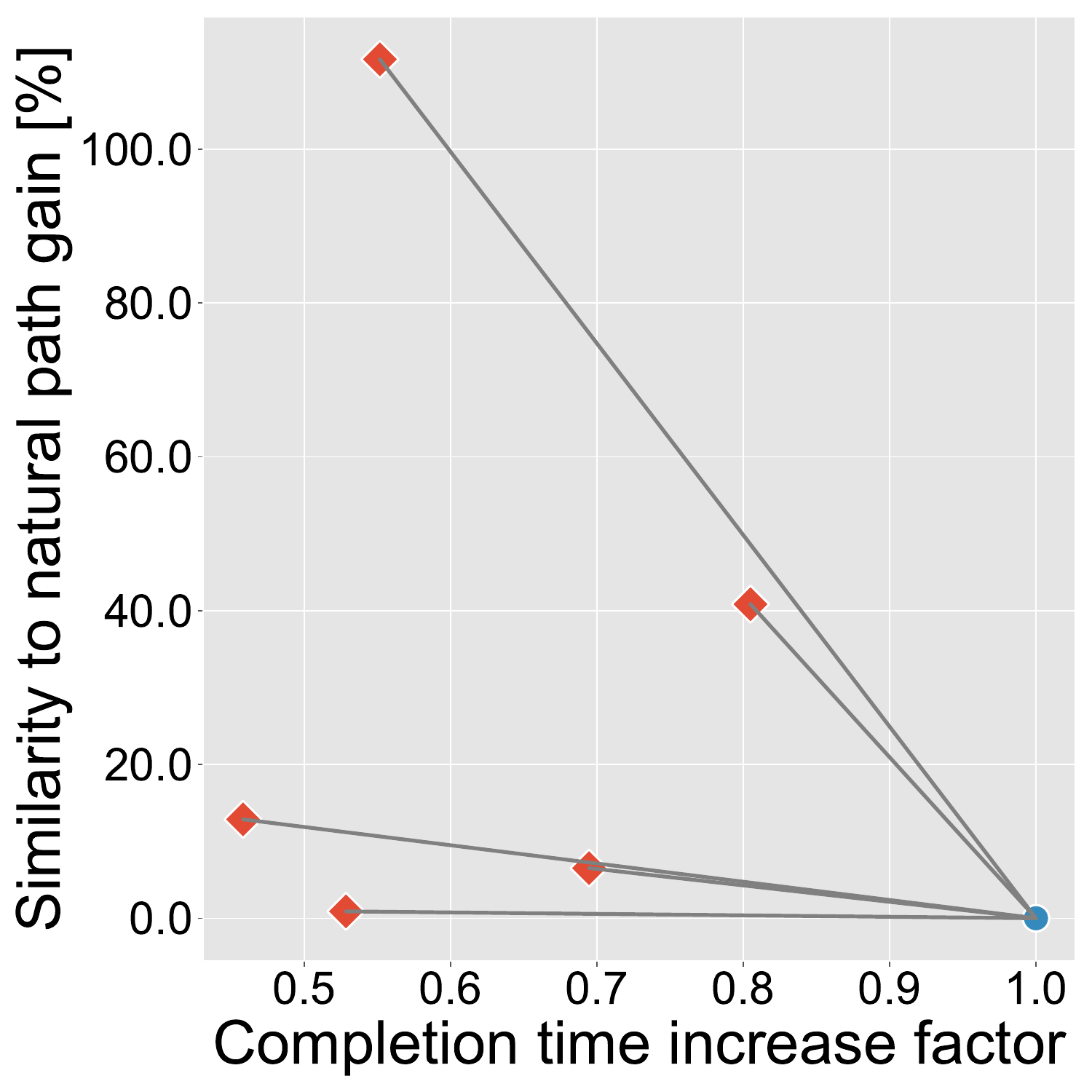} \\[-1mm]
    & (c) & (d) \\[1mm]
     &
    \includegraphics[width=0.45\widthForOneImage]{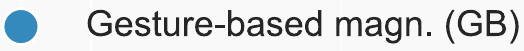} & 
    \includegraphics[width=0.45\widthForOneImage]{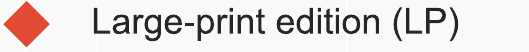} \\[-1mm]
    \end{tabular}
    \caption{\label{fig:phaseone}
    Average similarity to reference path vs average completion time per participant. 
    Absolute (left column) and relative (right column) performance plots for participants with normal vision (top row) and low vision (bottom row). Absolute plots (a,c) show the average similarity to the reference path as a function of average completion time per participant, for both \condonenot (blue circles) and \condtwonot (red diamonds). Relative plots (b,d) show the gains in similarity and the change in completion time when using \condtwonot (red diamonds) relative to \condonenot (blue circles).
    }
    \Description{This figure shows a 2-by-2 grid of four scatter plots comparing \condonenot and \condtwonot, on completion time (x-axis) and reading path similarity to the previously defined reference path (y-axis). The top row shows data for participants \NV, while the bottom row is for participants \LV. 

The left column (plots a, c) displays absolute values. In these plots, a line connects each participant's performance in the \condonenot condition (blue circle) to their performance in the \condtwonot condition (red diamond).

The right column (plots b, d) displays relative values, showing the change in performance when using \condtwonot relative to \condonenot, with the \condonenot baseline normalized to the point $(0,1)$.

Overall, the plots show that for both groups, \condtwonot was faster (red diamonds are positioned to the left of their corresponding blue circles) and more similar to the reference path (red diamonds are positioned to the top of their corresponding blue circle) }
\end{figure}

To visualize individual performance, \figref{fig:phaseone} plots reading path similarity against completion time for each participant, separated by \CVAS to account for different data ranges. The absolute performance plots (a,c) confirm that \condtwonot resulted in shorter times and paths more similar to the reference compared to \condonenot. The relative improvement plots (b, d) make these performance gains explicit. The plots also highlight that the improvements in both metrics were substantially larger for participants \LV than for participants \NV.

\subsubsection{Reading and transition times}
\label{section:reading-transition}

Our analysis of reading time revealed four significant factors: 
\begin{itemize}

    \item {Condition}: \replaced[comment="Typo before"]{There is a significant difference (\statreport{1}{209.24}{6.02}{=.01}{0.05}) between \condtwonot and \condonenot \comparison{$4.18$\seconds}{$3.80$\seconds}. However, due to the small effect size, we cannot confirm that reading time is the primary factor influencing the difference in completion time. 
    Additionally, a significant effect in the interaction with NewspaperID (\statreport{5}{209.02}{4.97}{<.001}{0.11}) was found. Post-hoc analysis indicated that the reading speed advantage of \condtwonot was isolated to specific layouts (Layouts \#13 and \#18, both $p<.001$), while no significant differences were found in the others. This indicates that the benefit on reading time is not universal, but highly dependent on specific structural arrangements.}
    {A significant effect found only for participants \NV, where \condtwonot had 15\% shorter times than \condonenot \comparison{$2.89$\seconds}{$2.47$\seconds}. The lack of significance for participants \LV may be due to their small sample size (only $5$ participants). This finding provides additional support for H1.}

    \item {\CVAS:} Participants \LV had lower reading times (avg. $5.29$\seconds) than participants \NV (avg. $2.68$\seconds) \added{(\statreport{1}{18.3}{15.50}{<.001}{0.46})}.

    \item {Trial index:} We found a significant but minor fatigue effect, with participants' reading times increasing in later trials, indicated by a positive slope \added{$(0.05)$} for this factor in our model  \added{(\statreport{1}{209.13}{7.66}{=.006}{0.04})}.

    \item {Content version:} A minor but significant difference \added{(\statreport{2}{211.57}{3.72}{=.02}{0.03})} occurred between version \#1 (avg. $3.84$\seconds) and version \#2 (avg. $4.0$\seconds), likely due to subtle variations in content complexity.
\end{itemize}

Our analysis of transition time revealed four significant factors:
\begin{itemize}
    \item {Condition}: \added{There is a significant difference (\statreport{1}{210.09}{73.568}{<.001}{0.29}) between \condonenot and \condtwonot \comparison{$4.13$\seconds}{$1.83$\seconds}. This effect size confirms that the transition time is the primary factor influencing completion time. However, the significant interaction with \CVAS \added{(\statreport{1}{209.47}{113.82}{<.001}{0.35})}, and the corresponding post-hoc analysis showed that this } significant effect occurred only for participants \LV \added{($p <.001$)}, who had $57\%$ lower transition time in \condtwonot \comparison{$8.0$\seconds}{$3.47$\seconds}. \replacedimx{This reduction was an expected mechanical consequence of}{as it} eliminating time-consuming manual interaction adjustments and facilitated direct access of the layout, allowing for immediate transition between headlines. In contrast, no significant difference was found for participants \NV \added{($p = 0.75$)}, likely due to their high efficiency and device familiarity in both conditions. This result supports \hypperformance.

    \item {\CVAS:} As expected, there was a large performance gap between groups \added{(\statreport{1}{19.58}{71.60}{<.001}{0.80})}, with participants \NV (avg. $0.214$\seconds) transitioning with a $96\%$ shorter time than participants \LV (avg. $5.74$\seconds).

    \item {NewspaperID:} \added{There is a significant difference (\statreport{5}{211.23}{7.45}{<.001}{0.15}), driven by Layout \#13 which} \deleted{Layout \#13} was significantly slower in transition time (avg. $4.05$\seconds) than the others (avg. $2.76$\seconds), likely due to a unique layout with a central article surrounded by the others, which might make the reading task more difficult. In any case, the main effect of \condtwonot superiority held in this layout, as well as in all the others.

    \item \replaced[comment="Typo before"]{{Trial Index:} Unlike reading time, we found a significant (but minor) reduction in transition time across trials, indicated by a negative slope \added{$(-0.09)$} for this factor in our model  \added{(\statreport{1}{209.62}{4.38}{=.037}{0.02})}. This suggests a learning effect regarding the experimental setup: participants became more proficient as the experiment progressed, even as reading fatigue set in.}{Content Version: A significant difference was found between version \#1 (avg $3.26$\seconds) and \#3 (avg $2.83$\seconds), which we again attribute to variations in content complexity.}
\end{itemize}

\begin{figure}[htpb]
    \centering
    \begin{tabular}{ccc}
    & {\bf Reading time} & {\bf Transition time}\\[1mm]
    \raisebox{25pt}[0pt][0pt]{\rotatebox{90}{{\bf \nv}}} &
    \includegraphics[width=0.45\widthForOneImage]{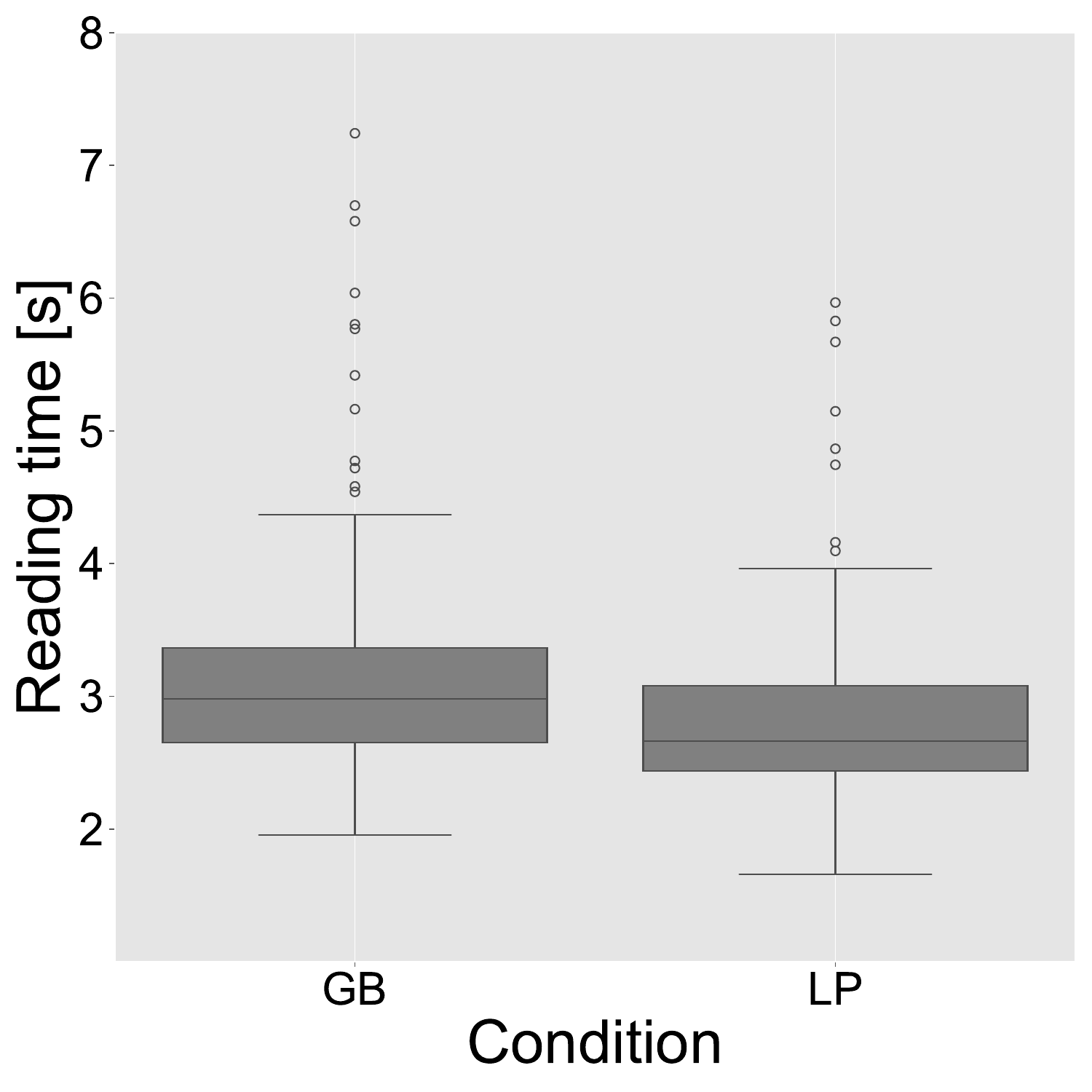} & 
    \includegraphics[width=0.45\widthForOneImage]{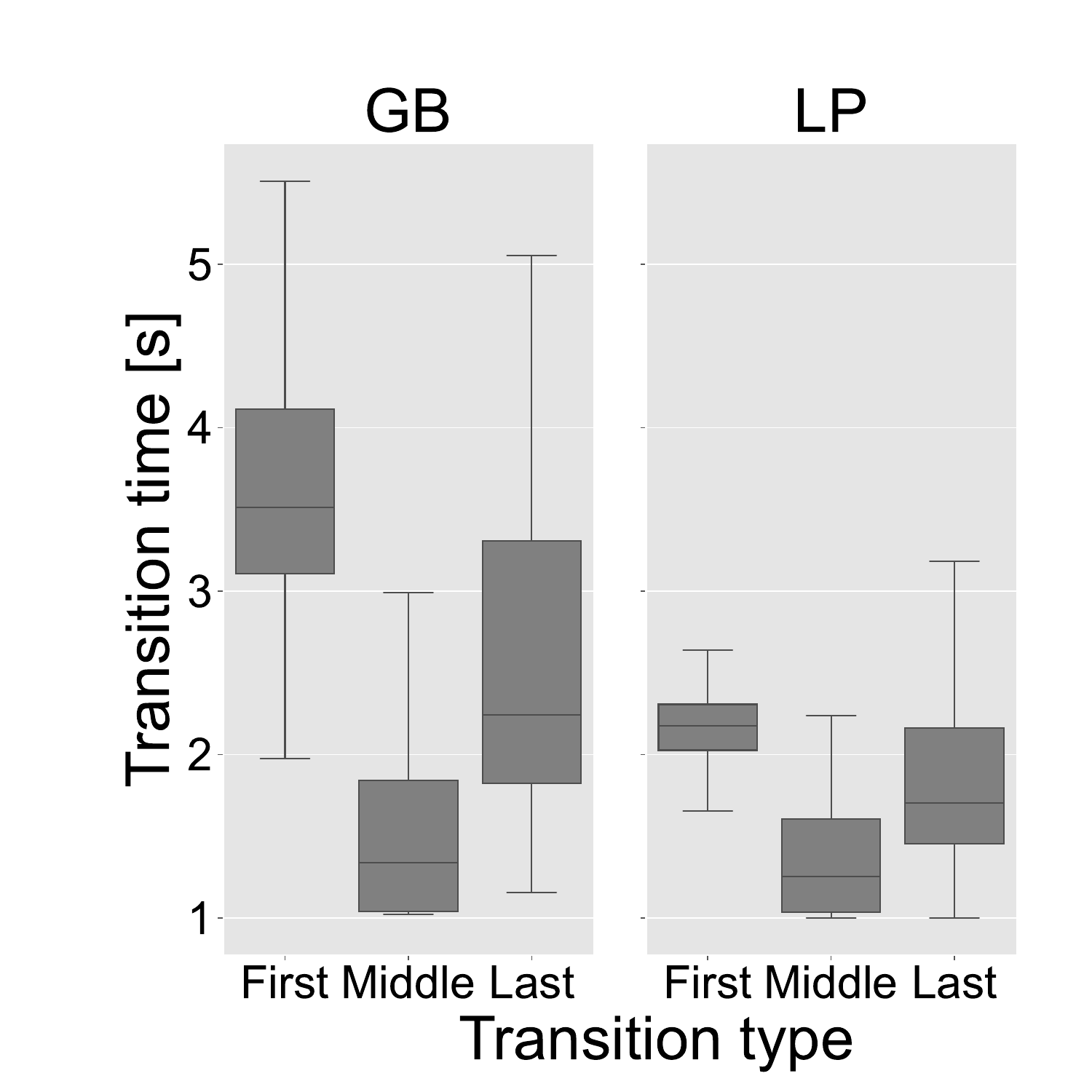} \\[-1mm]
    & (a) & (b) \\[1mm]
    \raisebox{25pt}[0pt][0pt]{\rotatebox{90}{{\bf \lv}}} &
    \includegraphics[width=0.45\widthForOneImage]{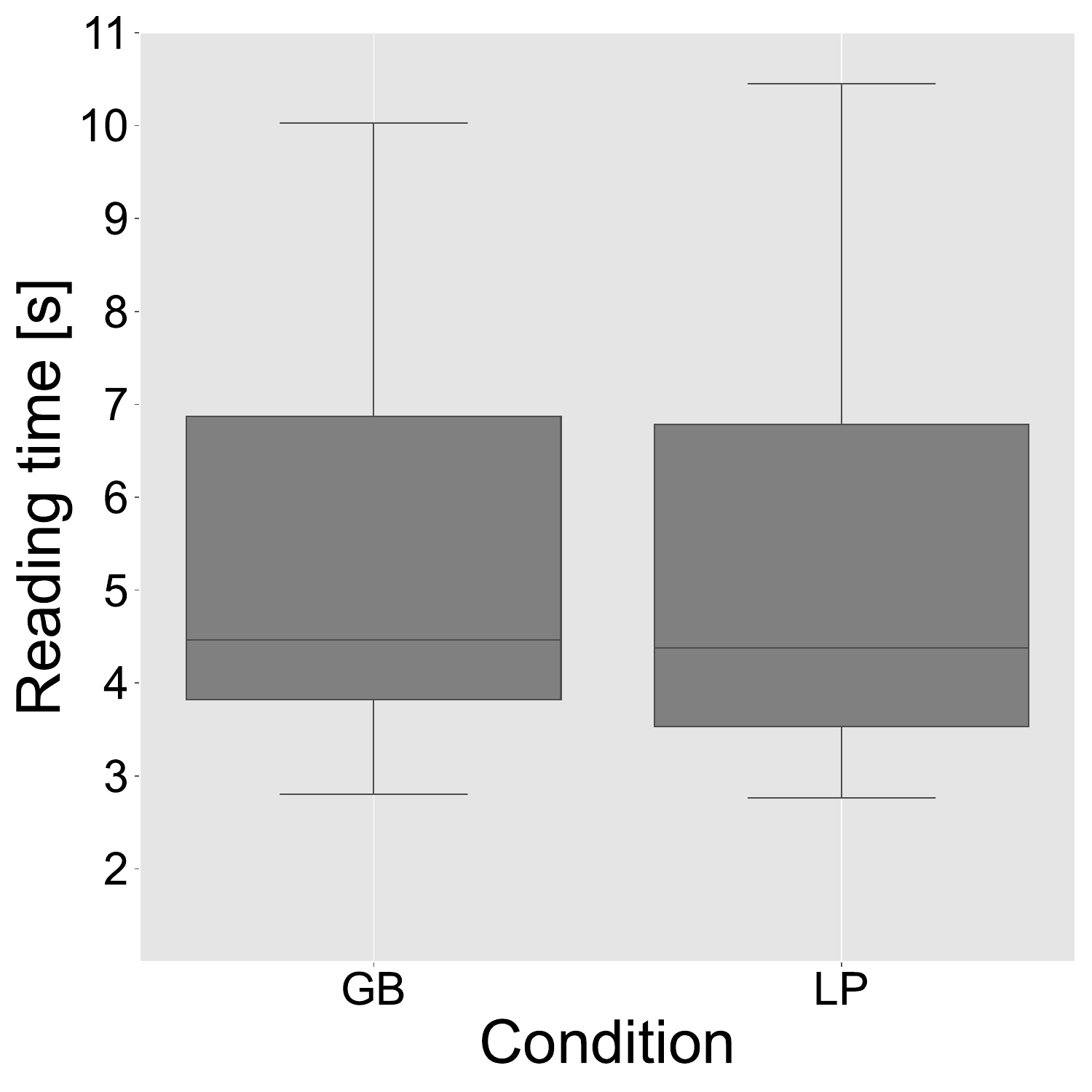} & 
    \includegraphics[width=0.45\widthForOneImage]{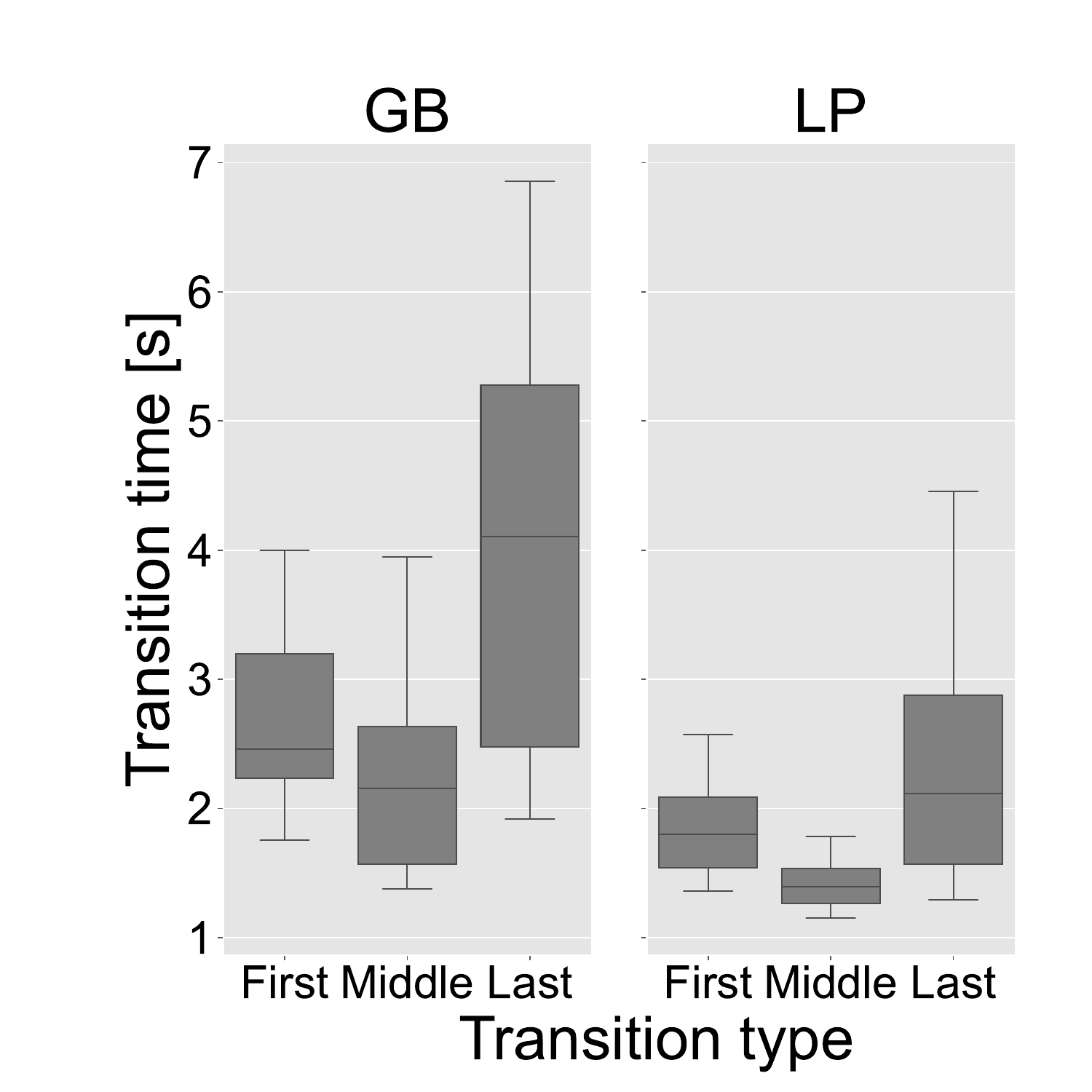} \\[-1mm]
    & (c) & (d)
    \end{tabular}
    \caption{\label{fig:phaseone-reading-transition}
    Reading and transition time per condition (Task 1). Reading time (left column) and transition time (right column) boxplots for participants \NV (top row) and \LV (bottom row).
    (a, c)  shows reading time boxplots. (b, d) shows transition time boxplot, separated by transition type: first transition (from stimulus onset to first headline), average of middle transitions (between headlines) and last transition (from last article to end of task). Outliers were removed to improve visualization}
    
    \Description{A 2-by-2 grid of box plots compares user performance in \condonenot and \condtwonot conditions. The top row shows data for participants \NV, and the bottom row for those \LV.

The left column (plots a, c) displays reading time. The difference is significant only for participants \NV (plot a).

The right column (plots b, d) displays transition time, broken down by 'First', 'Middle', and 'Last' transitions.

    For participants \NV, transition times are slightly faster but not significant in \condtwonot compared to the \condonenot

    For participants \LV, the difference is bigger and significant: transition times in \condtwonot are substantially faster than \condonenot.}
\end{figure}

Figure~\ref{fig:phaseone-reading-transition} visualizes the average reading and transition times for each participant, separated by \CVAS. The plots confirm that both reading times (a, c) and transition times (b, d) were shorter with \condtwonot than with \condonenot.

Moreover, the transition time plots (b, d) reveal that the difference is primarily driven by the first and last transitions. This reflects the initial setup time required in \condonenot for participants to adjust the zoom and pan to the first article. This effect was particularly pronounced for participants \LV, who were less familiar with touch gestures. These graphs underscore how layouts that eliminate manual interaction can significantly improve reading performance for users in \CVAS.

\subsection{Task 2 analysis}
Regarding Task 2, we derived two metrics to test hypotheses \hypfindingperformance and \hypfindingpositions:
\begin{enumerate}
    \item Success ratio: for this task, the success ratio was a binary measure (1 = target headline found, 0 = not found).
    \item Completion time: The total task duration, from stimulus onset to completion. 
\end{enumerate}

\subsubsection{Success ratio}

Our analysis revealed a significant effect of \added{NewspaperID (\statreport{5}{219.47}{4.10}{= 0.001}{0.09}), and an interaction between NewspaperID and \CVAS  (\statreport{5}{219.66}{ 3.30}{= .006}{0.07})} related to one specific newspaper layout (\#13). This was traced to a single participant \LV who, likely due to fatigue on his final trials, failed to find the target in either condition. Given the small sample size, this single failure disproportionately skewed the result for that specific case ($48.9\%$ success ratio).

Apart from this statistical artifact, no other factors had a significant effect on the success ratio. Overall, participants performed very well, achieving a success rate of over 90\% in both the \condonenot and \condtwonot conditions.

\subsubsection{Completion time}

 We extracted completion time for each trial, measured in seconds. Our analysis identified two significant factors:
\begin{itemize}
    \item {Condition:}  \added{There is a significant difference between conditions (\statreport{1}{212.63}{15.06}{<.001}{0.12})}. Overall, participants had $50\%$ shorter completion time with \condtwonot than with \condonenot \comparison{$18.7$\seconds}{$9.2$\seconds}. This effect was observed across both \CVAS \added{ as shown by the significance of the interaction (\statreport{1}{210.63}{14.81}{<.001}{0.07})}, with a $51\%$ improvement for participants \NV  \comparison{$6.91$\seconds}{$3.36$\seconds} and a $50\%$ improvement for participants \LV \comparison{$30.37$\seconds}{$15.03$\seconds}. \replacedimx{We attribute this to the navigational cost of manual interaction: in \condonenot, the effort required to manage the interaction distracts from the structural retrieval task. In contrast, \condtwonot allows for direct perception of the layout, reducing the time spent on interface management.} {We attribute this to the inherent ``control vs global awareness'' trade-off: in \condonenot, the lack of permanent global awareness forced participants to perform ``zooming in and out'' movements to verify headline locations, whereas \condtwonot provided immediate spatial context.} This result provides strong support for \hypfindingperformance.

     \item {\CVAS: }Participants \NV located the target considerably faster (avg. $5.14$\seconds) than participants \LV (avg. $22.7$\seconds) \added{(\statreport{1}{23.55}{47.20}{<.001}{0.71})}.
\end{itemize}

\begin{figure}[htpb]
    \centering
    \begin{tabular}{cc}
    {\bf \nv} & {\bf \lv}\\[1mm]
     
    \includegraphics[width=0.45\widthForOneImage]{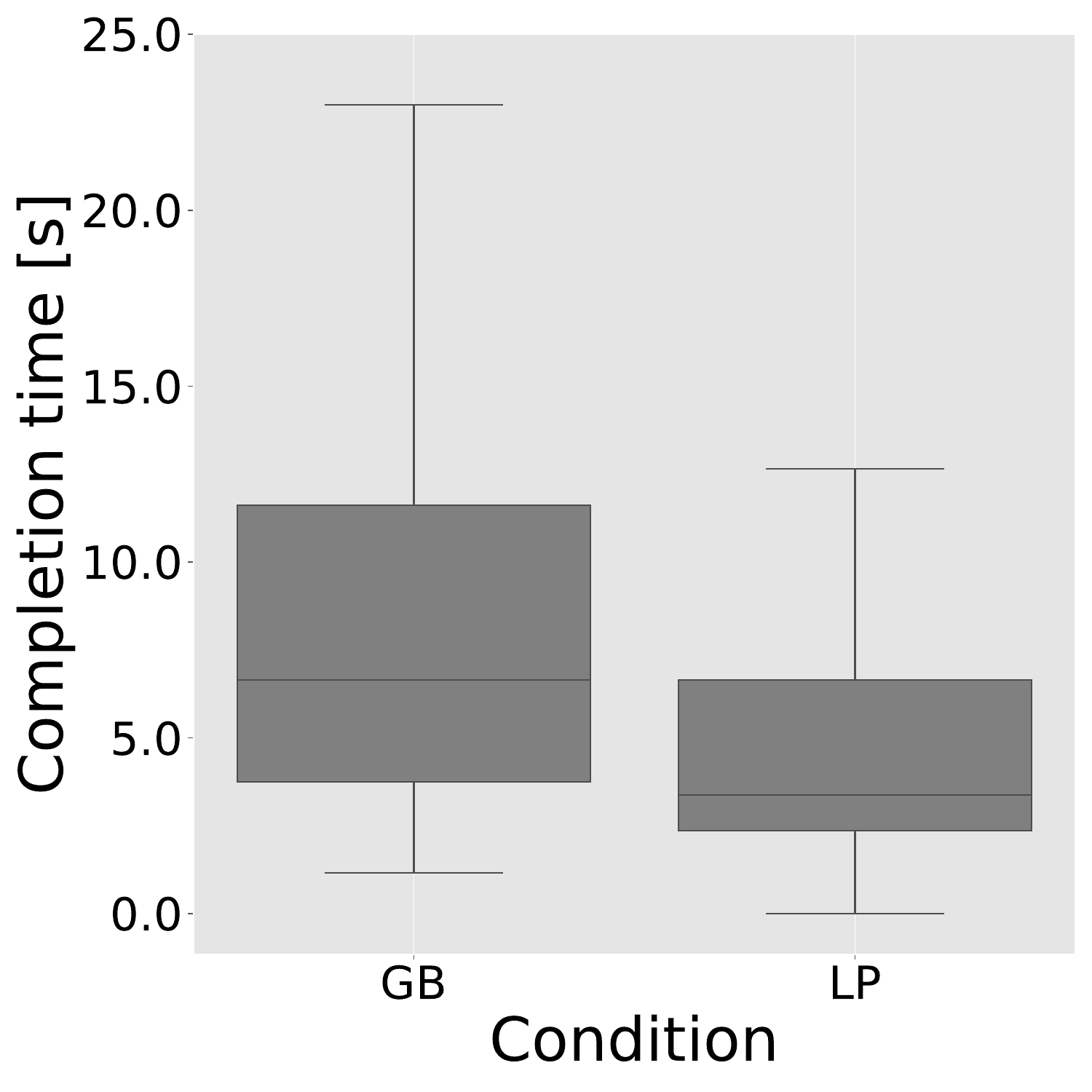} & 
    \includegraphics[width=0.45\widthForOneImage]{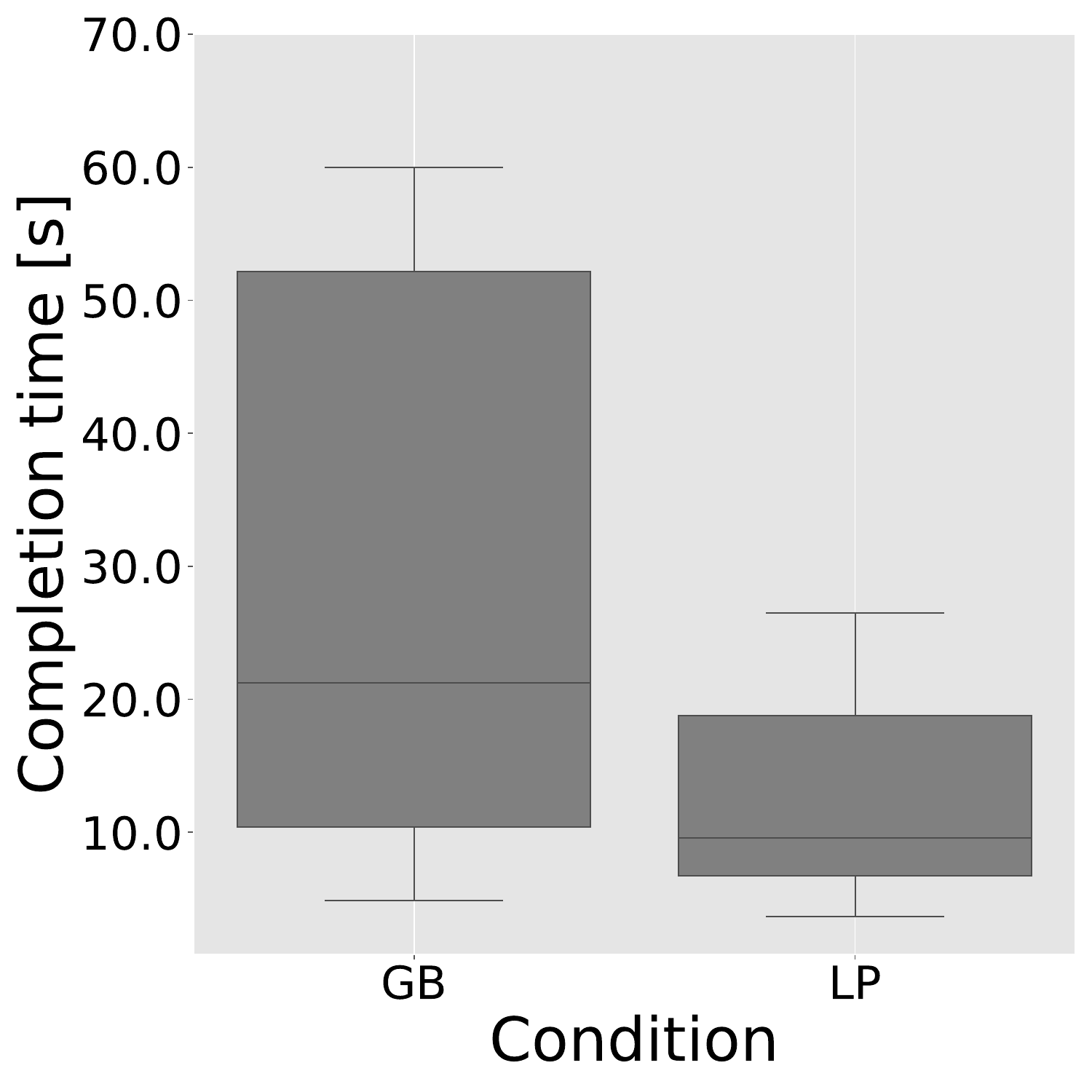} \\[-1mm]
     (a) & (b) \\[1mm]
    \end{tabular}
    \caption{\label{fig:phasetwo}
    Completion time per condition (Task 2), only successful cases.
    Boxplots showing participants’ distributions, with (a) normal vision and (b) low vision.}

    \Description{The figure displays two side-by-side box plots comparing Task 2 completion time between \condonenot and a \condtwonot.

    Plot (a) shows data for participants \NV. It indicates that \condtwonot was faster and had less variability (median time of approx. $3.5$ seconds) compared to \condonenot (median time of approx. $6$ seconds).

    Plot (b) shows data for participants \LV. The difference here is much more significant. \condtwonot was dramatically faster and more consistent (median time of approx. $9$ seconds) than \condonenot, which resulted in very high and widely variable completion times (median time of approx. $21$ seconds).}
\end{figure}

We can observe these results in \figref{fig:phasetwo}.

\subsubsection{Headline zone in Task 2}

Finally, we explored whether the target headline's location ("zone") influenced task completion time. \replaced[comment="Recency"]{The analysis revealed no statistically significant effect (\statreport{2}{209.75}{2.73}{=.067}{0.03})}{  We found a non-significant trend $(p<0.1)$ suggesting a possible recency effect; our recordings indicated that participants often recalled the location of headlines at the bottom of the page more easily. While this indicates a tendency, the result is not statistically significant,} and therefore we cannot confirm the hypothesis \hypfindingpositions.

\subsection{Participants preferences}
\label{subsection:preferences}

To test \hyppreference, we analyzed NASA-TLX scores, a direct preference question, and qualitative feedback.

\begin{figure}[htpb]
    \centering
    \begin{tabular}{cc}
    {\bf \nv} & {\bf \lv}\\[1mm]
     
    \includegraphics[width=\widthForTwoImages]{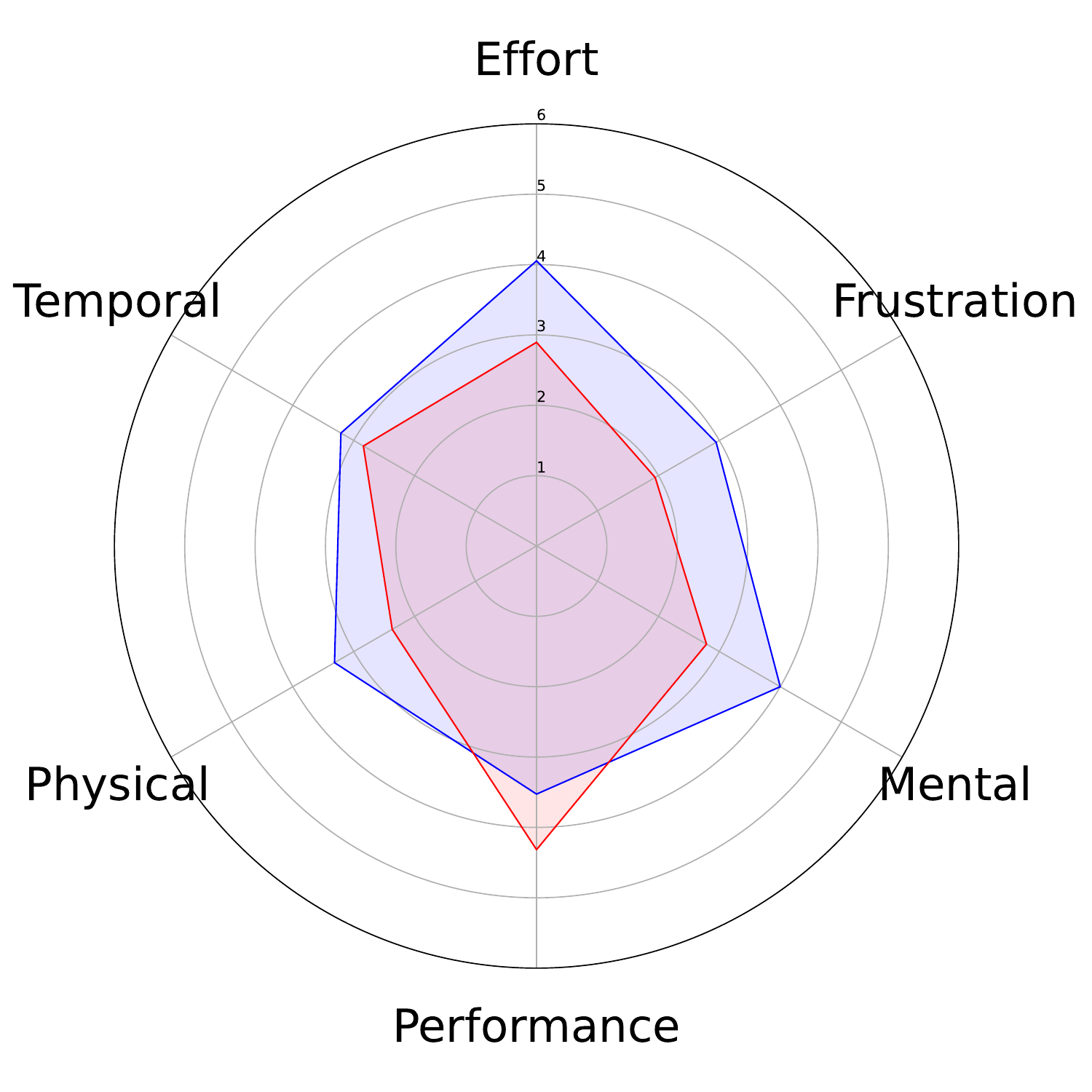} & 
    \includegraphics[width=\widthForTwoImages]{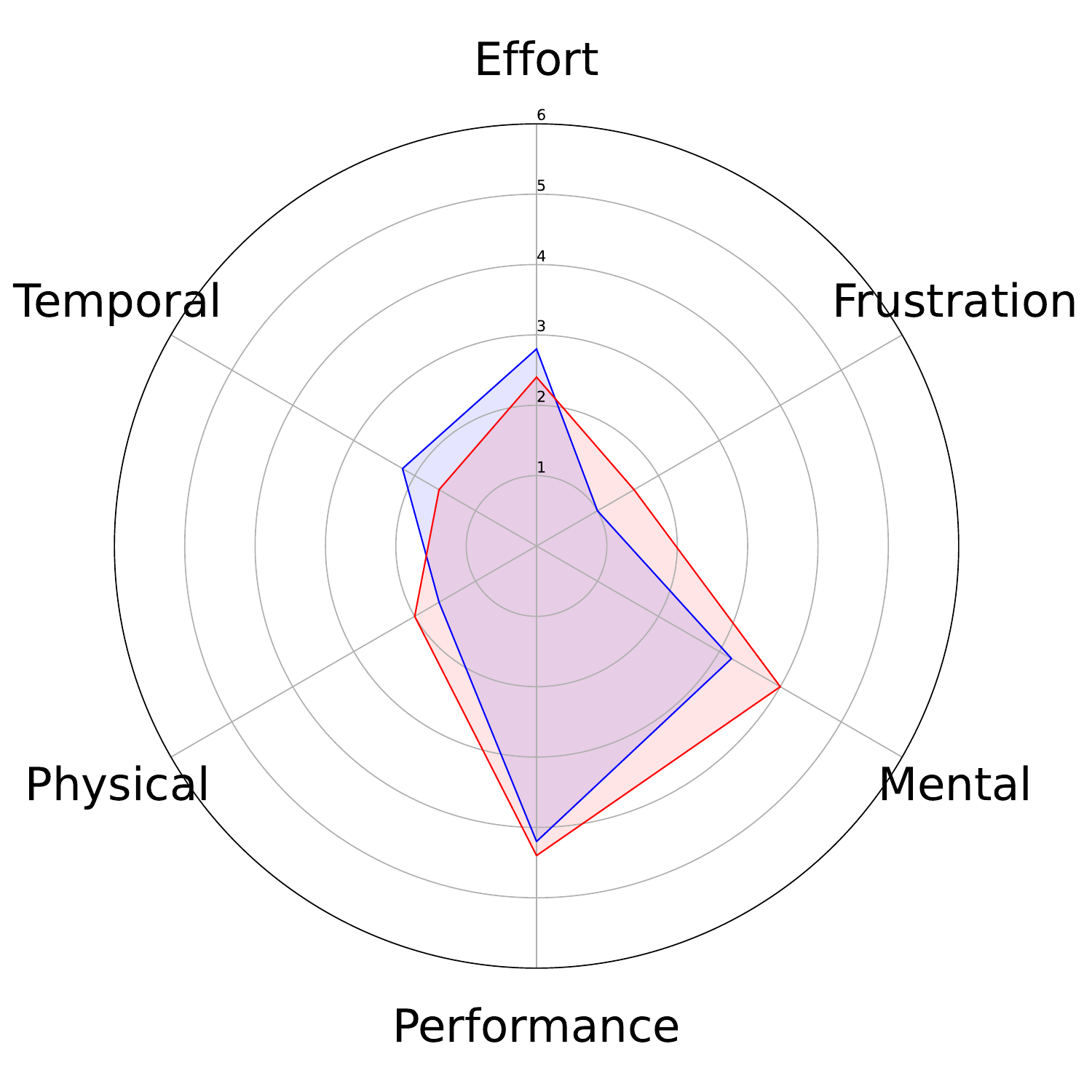} \\[-0.5mm]
     (a) & (b) \\[1mm]
    \includegraphics[width=0.2\widthForOneImage]{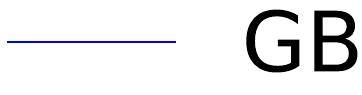} & 
    \includegraphics[width=0.2\widthForOneImage]{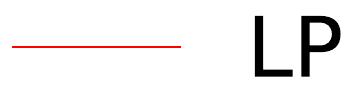} \\[-1mm]
    \end{tabular}
    \caption{\label{fig:nasa}
    NASA-TLX average scores for each question, for both \condonenot and \condtwonot, collected after completing the experiment. For all dimensions, lower scores indicate better perceived workload, except for performance, where higher scores are better.
(a) Participants \NV show a clear advantage for \condtwonot.
(b) Participants \LV show similar trends, though results should be interpreted cautiously due to the small sample size, which can skew the results, and higher subjectivity of their responses, as tasks were generally more demanding and ratings less consistent across conditions (see text for more details).}
    \Description{The figure contains two side-by-side radar charts that display average NASA-TLX workload scores, comparing \condonenot and \condtwonot.

Each chart has six axes: Effort, Frustration, Mental, Performance, Physical, and Temporal. For all axes except Performance, a lower score (closer to the center) is better. For the Performance axis, a higher score (further from the center) is better.

    Plot (a) shows results for participants \NV. The red polygon (\condtwonot) is consistently rated more favorably than the blue polygon (\condonenot), indicating \condtwonot required less workload and was perceived as having better performance.

    Plot (b) shows results for participants \LV. The trend is the same except for physical, mental and frustration.}
    
\end{figure}

NASA-TLX results (\figref{fig:nasa}) showed that participants \NV favored \condtwonot across all dimensions. In contrast, participants \LV reported higher physical and mental demand, together with more frustration for \condtwonot. Further analysis revealed this was driven by participants with the most severe visual impairments (i.e., higher CPS). \replaced[comment="5. ecological scenario."]{This finding requires careful contextualization: while producing \condtwonot based on CPS is ecologically valid, our strict requirement to render full headlines for experimental fairness between conditions imposed a magnification ceiling that frustrated these participants. In a real-world deployment, this could be mitigated by summarizing or truncating content to allow even higher magnification levels. However, for extreme visual constraints requiring extreme magnification levels, preserving a spatial 2D layout will inevitably become unfeasible, necessitating a shift to alternative solutions such as linear presentations (feed-like formats), as explored in prior work~\cite{almeraj_user_2019}.}{who may have required magnification levels beyond our algorithm's limits, which requires writing the entire headline in a limited space. We suggest that algorithmic improvements (e.g., summarizing text to make it shorter and "making room" for larger magnification factors) could address this and shift their preference.}

We triangulated these findings with objective performance (the sum of Task 1 and 2 completion times) and subjective preference. Based on total time, all participants ($100\%$) performed better with \condtwonot. The final preference question aligned with NASA-TLX results for most participants (75\%), though some discrepancies arose (25\%). These discrepancies are not surprising, as NASA-TLX is a non-comparative tool known to be sensitive to task order and subjective recall. This subjectivity is likely amplified in participants \LV, for whom the tasks were generally much more demanding, leading to less consistent ratings across conditions. To address this, we asked participants \LV at the end of the study to indicate, for each NASA-TLX dimension, which condition they preferred. This post-hoc comparative measure captured their subjective preference, giving in fact a preference for \condtwonot for all NASA-TLX dimensions except effort and frustration, where there is no preference for either condition. This reinforces the overall advantage of \condtwonot. Nevertheless, a dedicated comparative workload assessment would provide a more reliable metric in future studies. 

Finally, qualitative feedback provided additional insights:
\begin{enumerate}
    \item Comfort and Focus: Participants favored \condtwonot for a more comfortable and focused experience. Several participants \LV even changed their preference from \condonenot to \condtwonot over the course of the experiment.

    \item Layout constraints: Even if they overall prefer \condtwonot, they do not like to read headlines with too many lines and/or word breaks, which is something the relayouting method~\cite{gallardo} minimizes. This comment reinforces our idea of doing a relayouting over the original newspaper page while increasing font size and not just the latter, but also opens opportunities to improve this relayouting algorithm.

    \item Desire for control: A small minority ($3$ of $24$, or $12.5\%$, \added{which corresponds to the participants with the highest CPS}) preferred \condonenot despite their lower performance. \replacedimx{This finding highlights that a subset of users may prioritize the sense of control provided by manual interaction, even when it imposes a significantly higher behavioral and performance cost by disrupting the structural perception of the layout.}{This aligns with the trade-off discussed in the introduction, where some users prioritize the sense of control and predictability of manual interaction over global awareness.}
    \item Feed-like layouts: Participants enjoyed the vertical, feed-like layouts possible in \condtwonot. However, \added[comment="2. LP"]{ as argued in the Introduction,} our goal is to support the complex, aesthetic layouts of realistic newspapers, for which simple feed-like designs are not a scalable solution, especially when ads or images are present, since they cannot be shrunk.

\end{enumerate}

Taken together, these quantitative and qualitative results confirm our preference hypothesis \hyppreference.

\subsection{Summary}
Based on our previous results, let us make a summary focused in each hypothesis:
\begin{itemize}
    \item[\hypperformance]Supported. \condtwonot significantly improved mechanical efficiency, with faster completion times and more efficient transitions between articles compared to the navigational overhead of \condonenot.

    \item[\hyppath]Supported. We established a reference path based on \condzeronot trials, which followed a left-to-right, top-to-bottom reading order. We then measured how similar the reading paths in \condonenot and \condtwonot were to this reference path. In both \CVAS, \condtwonot paths were significantly more similar to the reference path than \condonenot paths. \replacedimx{This confirms that \condtwonot preserves the behavioral integrity of the reading process, whereas manual interaction forces a shift in the reader’s exploration strategy.}{indicating that \condtwonot has a smaller impact on the reference, i.e., unconstrained, reading order.}

    \item[\hypfindingperformance]Supported. Locating a specific target headline was faster using \condtwonot, \addedimx{demonstrating the advantage of direct structural access in eliminating the navigational overhead and cost of fragmented manual interaction.}

    \item[\hypfindingpositions]Not supported. While we observed a non-significant trend related to target position, our analysis did not find sufficient statistical evidence to support this hypothesis.

    \item[\hyppreference]Supported. NASA-TLX scores, a direct preference question, and qualitative feedback all converged to show a strong overall user preference for \condtwonot.
\end{itemize}

\subsection{\addedcr{Discussion}}

\subsubsection{\addedcr{ConditionOrder impact}}
\addedcr{As shown in Table~\ref{tab:lmm}, we model conditions using a within-subject factor, which we counterbalance  by systematic alternation to balance order (illustrated in Fig.~\ref{fig:conditions-setup}).
In addition, statistical analysis on ConditionOrder showed no significant results, indicating a lack of fatigue effects. }

\subsubsection{\addedcr{Generalization of the study}}

\addedcr{Our findings are expected to generalize within structured, layout-based documents, as loss of global context and interface manipulability are content-independent properties of GB vs. LP.
While this study quantified text-only reading, the methodology could potentially be extended to richer layouts with additional entry points~\cite{ozretic_dosen_key_2018}. This would require identifying the specific natural reading paths associated with such layouts, which would differ from the one defined here. Furthermore, we recognise that future protocols would likely need adaptations to address this use case, such as moving beyond reading aloud or accounting for CPS limitations when integrating visual elements.}

\subsubsection{\addedcr{Training and familiarity effects}}

\addedcr{The potential role of training and familiarity also merits consideration; observed patterns were similar for participants with normal-vision (already familiar with magnification) and low vision, though effects were larger for the latter. This suggests that training could reduce difference magnitude for them, but their visual acuity will prevent eliminating the gap, and the qualitative pattern between GB and LP is unlikely to change.}
\section{Conclusion} 
\label{sec:conclusion}

In this work, we quantified the behavioral and performance cost of consuming digital layout-based media when visual access is constrained. This challenge affects both readers on small-screen devices and people with low vision.

Our multidimensional study provides empirical evidence 
\replacedimx{and quantifies the impact of different newspaper interaction modalities. We found that while manual magnification (\condone) is the standard, it imposes a navigational overhead that disrupts the reader's canonical strategy. In contrast, \condtwo acts as a modality that restores behavioral integrity.}{regarding the trade-off between giving control over the magnification to the user and global awareness of the document. While \condone is the standard, we found that \condtwo offers superior outcomes. }
 Compared to gesture-based magnification, \condtwo enabled faster performance and a natural reading path (closely resembling the reference strategy), a stronger mental representation of document structure, and a significantly lower perceived workload.
 
These results indicate that magnification alone may be insufficient to support effective reading and consistent quality of experience in layout-based documents. The standard paradigm (\condone) can make navigating a page more challenging, not due to a lack of features, but by potentially disrupting spatial context and forcing a shift from natural content consumption. By contrast, re-layouting documents into large-print editions (\condtwo) preserves the spatial qualities of the layout (such as entry points) and supports efficient, predictable reading strategies.

Our findings offer important implications: designing \replacedimx{digital media consumption tools}{for accessibility} should go beyond interactive magnification tools toward adaptive layouts that combine font scaling with structural re-flow. Our findings confirm that the  effort required to generate \condtwo is justified \addedimx{not just} by the improvements in reader performance, \addedimx{but also by the measurable restoration of behavioral integrity} and satisfaction across diverse populations and devices, supporting efficient and inclusive access to newspapers, magazines, and other visually complex digital content. While our study focused on text-only layout-based documents, future work should extend these methods to richer layouts, incorporating graphical elements such as images and advertisements, and should include careful empirical studies to evaluate additional types of entry points.
\begin{acks}
This work was supported by the Association Nationale de la Recherche et de la Technologie (ANRT) through the CIFRE grant 2022/0927 in collaboration with Demain un Autre Jour.
\end{acks}

\bibliographystyle{ACM-Reference-Format}
\bibliography{references}

@book{lazar-hci-2017,
year = {2017},
author = {Lazar, Jonathan and Feng, Jinjuan Heidi and Hochheiser, Harry},
address = {Cambridge, Massachusetts},
booktitle = {Research methods in human-computer interaction},
edition = {Second edition},
isbn = {9780128093436},
language = {eng},
publisher = {Morgan Kaufmann Publishers, an imprint of Elsevier},
series = {Gale eBooks},
title = {Research methods in human-computer interaction },
}

@inproceedings{bujic-imx2023,
author = {Buji\'{c}, Mila and Salminen, Mikko and Hamari, Juho},
title = {More Immersed but Less Present: Unpacking Factors of Presence Across Devices},
year = {2023},
isbn = {9798400700286},
publisher = {Association for Computing Machinery},
address = {New York, NY, USA},
url = {https://doi.org/10.1145/3573381.3596152},
doi = {10.1145/3573381.3596152},
booktitle = {Proceedings of the 2023 ACM International Conference on Interactive Media Experiences},
pages = {140–149},
numpages = {10},
location = {Nantes, France},
series = {IMX '23}
}

@inproceedings{vatavu-imx2021,
author = {Vatavu, Radu-Daniel},
title = {Accessibility of Interactive Television and Media Experiences: Users with Disabilities Have Been Little Voiced at IMX and TVX},
year = {2021},
isbn = {9781450383899},
publisher = {Association for Computing Machinery},
address = {New York, NY, USA},
url = {https://doi.org/10.1145/3452918.3465485},
doi = {10.1145/3452918.3465485},
booktitle = {Proceedings of the 2021 ACM International Conference on Interactive Media Experiences},
pages = {218–222},
numpages = {5},
location = {Virtual Event, USA},
series = {IMX '21}
}

@misc{marcotte_responsive_2010,
  author       = {Marcotte, Ethan},
  title        = {Responsive Web Design},
  howpublished = {\url{https://alistapart.com/article/responsive-web-design/}},
  year         = {2010},
  note         = {Accessed: December 3rd, 2025},
  publisher    = {A List Apart}
}

@mastersthesis{chesham_master_2003,
  author  = {Chesham, Andrew P.},
  title   = {Poetry and eBooks: Managing Fixed Layouts in Reflowable Text},
  school  = {University of Simon Fraser},
  year    = {2003},
  type    = {Master's Thesis},
}

@inproceedings{chiou_2024,
	title = {Automatically {Detecting} {Reflow} {Accessibility} {Issues} in {Responsive} {Web} {Pages}},
	isbn = {9798400702174},
	url = {https://dl.acm.org/doi/10.1145/3597503.3639229},
	doi = {10.1145/3597503.3639229},
	language = {en},
	urldate = {2025-12-03},
	booktitle = {Proceedings of the {IEEE}/{ACM} 46th {International} {Conference} on {Software} {Engineering}},
	publisher = {ACM},
    address = {New York, NY, USA},
	author = {Chiou, Paul T. and Winn, Robert and Alotaibi, Ali S. and Halfond, William G. J.},
	month = apr,
	year = {2024},
	pages = {1--13}
}

@inproceedings{ohara_comparison_1997,
author = {O'Hara, Kenton and Sellen, Abigail},
title = {A comparison of reading paper and on-line documents},
year = {1997},
isbn = {0897918029},
publisher = {Association for Computing Machinery},
address = {New York, NY, USA},
url = {https://doi.org/10.1145/258549.258787},
doi = {10.1145/258549.258787},
booktitle = {Proceedings of the ACM SIGCHI Conference on Human Factors in Computing Systems},
pages = {335–342},
numpages = {8},
location = {Atlanta, Georgia, USA},
series = {CHI '97}
}

@inproceedings{almeraj_user_2019,
author = {Almeraj, Zainab and Alsumait, Asmaa},
year = {2019},
month = {02},
pages = {},
title = {A User Centered Design Roadmap for Researchers and Designers Working with Visually Impaired and Blind Children},
booktitle = {ACHI 2019 : The Twelfth International Conference on Advances in Computer-Human Interactions},
publisher = {IARIA},
address = {Athens, Greece}
}

@inproceedings{wang_gazeprompt_2024,
	address = {Honolulu HI USA},
	title = {{GazePrompt}: {Enhancing} {Low} {Vision} {People}'s {Reading} {Experience} with {Gaze}-{Aware} {Augmentations}},
	isbn = {9798400703300},
	shorttitle = {{GazePrompt}},
	url = {https://dl.acm.org/doi/10.1145/3613904.3642878},
	doi = {10.1145/3613904.3642878},
	language = {en},
	urldate = {2025-11-19},
	booktitle = {Proceedings of the {CHI} {Conference} on {Human} {Factors} in {Computing} {Systems}},
	publisher = {ACM},
	author = {Wang, Ru and Potter, Zach and Ho, Yun and Killough, Daniel and Zeng, Linxiu and Mondal, Sanbrita and Zhao, Yuhang},
	month = may,
	year = {2024},
	pages = {1--17},
}

@inproceedings{wang_understanding_2023,
	address = {Hamburg Germany},
	title = {Understanding {How} {Low} {Vision} {People} {Read} {Using} {Eye} {Tracking}},
	isbn = {978-1-4503-9421-5},
	url = {https://dl.acm.org/doi/10.1145/3544548.3581213},
	doi = {10.1145/3544548.3581213},
	language = {en},
	urldate = {2025-11-19},
	booktitle = {Proceedings of the 2023 {CHI} {Conference} on {Human} {Factors} in {Computing} {Systems}},
	publisher = {ACM},
	author = {Wang, Ru and Zeng, Linxiu and Zhang, Xinyong and Mondal, Sanbrita and Zhao, Yuhang},
	month = apr,
	year = {2023},
	pages = {1--17},
}

@article{valsecchi_saccadic_2013,
	title = {Saccadic and smooth-pursuit eye movements during reading of drifting texts},
	volume = {13},
	issn = {1534-7362},
	url = {http://jov.arvojournals.org/Article.aspx?doi=10.1167/13.10.8},
	doi = {10.1167/13.10.8},
	language = {en},
	number = {10},
	urldate = {2025-11-19},
	journal = {Journal of Vision},
	author = {Valsecchi, M. and Gegenfurtner, K. R. and Schutz, A. C.},
	month = aug,
	year = {2013},
	pages = {8--8},
}

@incollection{jacob_eye_2003,
	title = {Eye Tracking in Human-Computer Interaction and Usability Research: Ready to Deliver the Promises},
    editor = {J. Hyönä and R. Radach and H. Deubel},
    booktitle = {The Mind's Eye},
    publisher = {North-Holland},
    address = {Amsterdam},
    pages = {573-605},
    year = {2003},
    isbn = {978-0-444-51020-4},
    doi = {https://doi.org/10.1016/B978-044451020-4/50031-1},
    url = {https://www.sciencedirect.com/science/article/pii/B9780444510204500311},
    author = {Robert J.K. Jacob and Keith S. Karn}
}

@inproceedings{heo_reading_2024,
	address = {Glasgow United Kingdom},
	title = {Reading with {Screen} {Magnification}: {Eye} {Movement} {Analysis} {Using} {Compensated} {Gaze} {Tracks}},
	isbn = {9798400706073},
	shorttitle = {Reading with {Screen} {Magnification}},
	url = {https://dl.acm.org/doi/10.1145/3649902.3656493},
	doi = {10.1145/3649902.3656493},
	language = {en},
	urldate = {2025-11-19},
	booktitle = {Proceedings of the 2024 {Symposium} on {Eye} {Tracking} {Research} and {Applications}},
	publisher = {ACM},
	author = {Heo, Seongsil and Manduchi, Roberto and Chung, Suzana},
	month = jun,
	year = {2024},
	pages = {1--6},
}

@article{ozretic_dosen_key_2018,
	title = {Key design elements of daily newspapers: {Impact} on the reader's perception and visual impression},
	issn = {20637330},
	shorttitle = {Key design elements of daily newspapers},
	url = {http://komejournal.com/files/KOME_DOD&LB.pdf},
	doi = {10.17646/KOME.75692.93},
	language = {en},
	urldate = {2025-11-19},
	journal = {KOME},
	author = {Ozretić Došen, Đurđana and Brkljačić, Lidija},
	month = aug,
	year = {2018},
}

@article{hollander_e-reader_2011,
	title = {The {E}-{Reader} as {Replacement} for the {Print} {Newspaper}},
	volume = {27},
	doi = {10.1007/s12109-011-9205-8},
	journal = {Pub Res Q},
	author = {Hollander, Barry A. and Krugman, Dean M. and Reichert, Tom and Avant, J. Adam},
	year = {2011},
	pages = {126--134},
}

@inproceedings{tang_screen_2023,
	address = {New York NY USA},
	title = {Screen {Magnification} for {Readers} with {Low} {Vision}: {A} {Study} on {Usability} and {Performance}},
	isbn = {9798400702204},
	shorttitle = {Screen {Magnification} for {Readers} with {Low} {Vision}},
	url = {https://dl.acm.org/doi/10.1145/3597638.3608383},
	doi = {10.1145/3597638.3608383},
	language = {en},
	urldate = {2025-05-16},
	booktitle = {The 25th {International} {ACM} {SIGACCESS} {Conference} on {Computers} and {Accessibility}},
	publisher = {ACM},
	author = {Tang, Meini and Manduchi, Roberto and Chung, Susana and Prado, Raquel},
	month = oct,
	year = {2023},
	pages = {1--15},
}

@phdthesis{buring_zoomable,
	title = {Zoomable {User} {Interfaces} on {Small} {Screens} - {Presentation}  {Interaction} {Design} for {Pen}-{Operated} {Mobile} {Devices}},
	author = {Büring, Thorsten},
	year = 2001,
	month = {July},
	type = {PhD Thesis},
	school = {University of Konstanz},

}

@article{atata_evaluating_2025,
	title = {Evaluating {Visual} {Hierarchies} and {Navigation} {Patterns} to {Improve} {Accessibility}},
	volume = {6},
	issn = {25827421},
	url = {https://ijarpr.com/uploads/V2ISSUE8/IJARPR0851.pdf},
	doi = {10.55248/gengpi.6.0825.3080},
	language = {en},
	number = {8},
	urldate = {2025-11-19},
	journal = {International Journal of Research Publication and Reviews},
	author = {Atata, Raheemat},
	month = aug,
	year = {2025},
	pages = {617--641},
}

@mastersthesis{aljoudi,
	title = {Direct {Manipulation} of {AI} {Text} via {Gestures}},
	language = {en},
	year = {2020},
	author = {Aljoudi, Ahmad},
    school = {Malmö University, Faculty of Culture and Society (KS), School of Arts and Communication (K3)},
	type={Bachelor's Thesis}
}

@article{priyadarshini_impact_2024,
	title = {The {Impact} of {User} {Interface} {Design} on {User} {Engagement}},
	volume = {13},
	language = {en},
	number = {03},
	journal = {International Journal of Engineering Research},
	author = {Priyadarshini, Ar Poorva},
	year = {2024}
}

@article{shneiderman_direct_1983,
	title = {Direct {Manipulation}: {A} {Step} {Beyond} {Programming} {Languages}},
	volume = {16},
	copyright = {https://ieeexplore.ieee.org/Xplorehelp/downloads/license-information/IEEE.html},
	issn = {0018-9162},
	shorttitle = {Direct {Manipulation}},
	url = {http://ieeexplore.ieee.org/document/1654471/},
	doi = {10.1109/MC.1983.1654471},
	language = {en},
	number = {8},
	urldate = {2025-11-19},
	journal = {Computer},
	author = {{Shneiderman}},
	month = aug,
	year = {1983},
	
pages = {57--69},
}

@article{granquist_how_2018,
	title = {How {People} with {Low} {Vision} {Achieve} {Magnification} in {Digital} {Reading}},
	volume = {95},
	issn = {1538-9235, 1040-5488},
	url = {https://journals.lww.com/00006324-201809000-00005},
	doi = {10.1097/OPX.0000000000001261},
	language = {en},
	number = {9},
	urldate = {2024-06-06},
	journal = {Optometry and Vision Science},
	author = {Granquist, Christina and Wu, Yueh-Hsun and Gage, Rachel and Crossland, Michael D. and Legge, Gordon E.},
	month = sep,
	year = {2018},
	pages = {711--719},
}

@misc{
whisperX,
      title={WhisperX: Time-Accurate Speech Transcription of Long-Form Audio}, 
      author={Max Bain and Jaesung Huh and Tengda Han and Andrew Zisserman},
      year={2023},
      eprint={2303.00747},
      archivePrefix={arXiv},
      primaryClass={cs.SD},
      url={https://arxiv.org/abs/2303.00747}, 
}

@misc{mistralai,
  title        = {MistralAI},
  author       = {Mistral AI},
  year         = 2025,
  note         = {Accessed: September 6th, 2025},
  howpublished = {\url{https://mistral.ai/}}
}

@misc{R,
  title        = {R: The R Project for Statistical Computing},
  author       = {R Project},
  year         = 2025,
  note         = {Accessed: September 6th, 2025},
  howpublished = {\url{https://www.r-project.org/}}
}

@article{legge1992psychophysics,
  author    = {Gordon E. Legge and John A. Ross and L. Michael Isenberg and J. M. LaMay},
  title     = {Psychophysics of Reading—Clinical Predictors of Low-Vision Reading Speed},
  journal   = {Investigative Ophthalmology \& Visual Science},
  volume    = {33},
  number    = {3},
  pages     = {677--687},
  year      = {1992}
}

@ARTICLE{levenshtein2007,
  author={Yujian, Li and Bo, Liu},
  journal={IEEE Transactions on Pattern Analysis and Machine Intelligence}, 
  title={A Normalized Levenshtein Distance Metric}, 
  year={2007},
  volume={29},
  number={6},
  pages={1091-1095},
  doi={10.1109/TPAMI.2007.1078}}

@article{readinghabits2016,
    title = {Why read it on your mobile device? Change in reading habit of electronic magazines for university students},
    journal = {The Journal of Academic Librarianship},
    volume = {42},
    number = {6},
    pages = {664-669},
    year = {2016},
    issn = {0099-1333},
    doi = {https://doi.org/10.1016/j.acalib.2016.08.007},
    url = {https://www.sciencedirect.com/science/article/pii/S0099133316301768},
    author = {Peng Wang and Dickson K.W. Chiu and Kevin K.W. Ho and Patrick Lo},
}

@article{mobilereading2015,
  title={An Overview of Mobile Reading Habits},
  author={Shimray, Somipam R. and Keerti, Chennupati and Ramaiah, Chennupati K.},
  journal={DESIDOC Journal of Library \& Information Technology},
  volume={35},
  number={5},
  pages={343--354},
  year={2015}
}

@article{reading-periodicals2021,
author = {Shao Jing Ding and Ernest Tak Hei Lam and Dickson KW Chiu and Mavis Man-wai Lung and Kevin KW Ho},
title ={Changes in reading behaviour of periodicals on mobile devices: A comparative study},
journal = {Journal of Librarianship and Information Science},
volume = {53},
number = {2},
pages = {233-244},
year = {2021},
doi = {10.1177/0961000620938119},
URL = { 
        https://doi.org/10.1177/0961000620938119
},
eprint = { 
        https://doi.org/10.1177/0961000620938119
}
}

@inproceedings{mansfield1993,
    author = {J. Stephen Mansfield and Sonia J. Ahn and Gordon E. Legge and Andrew Luebker},
    booktitle = {Noninvasive Assessment of the Visual System},
    journal = {Noninvasive Assessment of the Visual System},
    pages = {NSuD.3},
    publisher = {Optica Publishing Group},
    title = {Commentary on Section 4 - A New Reading-Acuity Chart for Normal and Low Vision},
    year = {1993},
    url = {https://opg.optica.org/abstract.cfm?URI=NAVS-1993-NSuD.3},
    doi = {10.1364/NAVS.1993.NSuD.3}
}

@unpublished{gallardo,
  TITLE = {{Newspaper Magnification with Preserved Entry Points}},
  AUTHOR = {Gallardo, Sebastian and Riff, Mar{\'i}a Cristina and Mazauric, Dorian and Kornprobst, Pierre},
  URL = {https://hal.science/hal-04210840},
  NOTE = {working paper or preprint},
  YEAR = {2023},
  MONTH = Sep,
  HAL_ID = {hal-04210840}
}

@article{atilgan2020,
author = {Atilgan, Nilsu and Xiong, Ying-Zi and Legge, Gordon E.},
title = {Reconciling print-size and display-size constraints on reading},
journal = {Proceedings of the National Academy of Sciences},
volume = {117},
number = {48},
pages = {30276-30284},
year = {2020},
doi = {10.1073/pnas.2007514117}
}

@article{Hajek2023,
author = {Hajek, André and Gyasi, Razak and Kretzler, Benedikt and König, Hans-Helmut},
year = {2023},
month = {11},
pages = {1-8},
title = {Vision and hearing problems and psychosocial outcomes: longitudinal evidence from the German Ageing Survey},
volume = {59},
journal = {Social Psychiatry and Psychiatric Epidemiology},
doi = {10.1007/s00127-023-02588-9}
}

@INPROCEEDINGS{shneiderman1996,
  author={Shneiderman, B.},
  booktitle={Proceedings 1996 IEEE Symposium on Visual Languages}, 
  title={The eyes have it: a task by data type taxonomy for information visualizations}, 
  year={1996},
  pages={336-343},
  doi={10.1109/VL.1996.545307}}

@inproceedings{svi-bright-env,
author = {Nikitenko, Denis and Evans, Jordan and Flatla, David R. and Driscoll, Thomas and Quinlan, Graham and Lukaszek, Kyle},
title = {Situational visual impairments on mobile devices - modeling the effects of bright outdoor environments},
year = {2024},
isbn = {9798400718281},
publisher = {Association for Computing Machinery},
address = {New York, NY, USA},
url = {https://doi.org/10.1145/3670947.3670959},
doi = {10.1145/3670947.3670959},
booktitle = {Proceedings of the 50th Graphics Interface Conference},
articleno = {1},
numpages = {10},
location = {Halifax, NS, Canada},
series = {GI '24}
}

@inproceedings{svi-challenges,
author = {Sarsenbayeva, Zhanna and van Berkel, Niels and Luo, Chu and Kostakos, Vassilis and Goncalves, Jorge},
title = {Challenges of situational impairments during interaction with mobile devices},
year = {2017},
isbn = {9781450353793},
publisher = {Association for Computing Machinery},
address = {New York, NY, USA},
url = {https://doi.org/10.1145/3152771.3156161},
doi = {10.1145/3152771.3156161},
booktitle = {Proceedings of the 29th Australian Conference on Computer-Human Interaction},
pages = {477–481},
numpages = {5},
location = {Brisbane, Queensland, Australia},
series = {OzCHI '17}
}

@inproceedings{tigwell-2018,
author = {Tigwell, Garreth W. and Menzies, Rachel and Flatla, David R.},
title = {Designing for Situational Visual Impairments: Supporting Early-Career Designers of Mobile Content},
year = {2018},
isbn = {9781450351980},
publisher = {Association for Computing Machinery},
address = {New York, NY, USA},
url = {https://doi.org/10.1145/3196709.3196760},
doi = {10.1145/3196709.3196760},
booktitle = {Proceedings of the 2018 Designing Interactive Systems Conference},
pages = {387–399},
numpages = {13},
location = {Hong Kong, China},
series = {DIS '18}
}

@misc{moran2016reading,
  author       = {Kate Moran},
  title        = {Reading Content on Mobile Devices},
  howpublished = {\url{https://www.nngroup.com/articles/mobile-content/}},
  note         = {Accessed: August 18th, 2025},
  year         = {2016},
}

@misc{Nielsen2011,
  author       = {Jakob Nielsen},
  title        = {Mobile Content Is Twice as Difficult},
  howpublished = {\url{https://www.nngroup.com/articles/mobile-content-is-twice-as-difficult-2011/}},
  note         = {Accessed: August 25th, 2025},
  year         = {2011},
}

@misc{Budiu2015,
  author       = {Raluca Budiu},
  title        = {Mobile User Experience: Limitations and Strengths},
  howpublished = {\url{https://www.nngroup.com/articles/mobile-ux/}},
  note         = {Accessed: August 25th, 2025},
  year         = {2015},
}

@unpublished{yue2024situfont,
  title        = {SituFont: A Just‑in‑Time Adaptive Intervention System for Enhancing Mobile Readability in Situational Visual Impairments},
  author       = {Kun Yue and Mingshan Zhang and Jingruo Chen and Chun Yu and Kexin Nie and Zhiqi Gao and Jinghan Yang and Chen Liang and Yuanchun Shi},
  year         = {2024},
  month        = oct,
  day          = {12},
  journal      = {arXiv preprint},
  archivePrefix= {arXiv},
  eprint       = {2410.09562},
  url          = {https://arxiv.org/abs/2410.09562},
}

@article{nav-patterns-2002,
author = {Hornb\ae{}k, Kasper and Bederson, Benjamin B. and Plaisant, Catherine},
title = {Navigation patterns and usability of zoomable user interfaces with and without an overview},
year = {2002},
issue_date = {December 2002},
publisher = {Association for Computing Machinery},
address = {New York, NY, USA},
volume = {9},
number = {4},
issn = {1073-0516},
url = {https://doi.org/10.1145/586081.586086},
doi = {10.1145/586081.586086},
journal = {ACM Trans. Comput.-Hum. Interact.},
month = dec,
pages = {362–389},
numpages = {28}
}

@article{aguilar-castet2017,
    doi = {10.1371/journal.pone.0174910},
    author = {Aguilar, Carlos AND Castet, Eric},
    journal = {PLOS ONE},
    publisher = {Public Library of Science},
    title = {Evaluation of a gaze-controlled vision enhancement system for reading in visually impaired people},
    year = {2017},
    month = {04},
    volume = {12},
    url = {https://doi.org/10.1371/journal.pone.0174910},
    pages = {1-24},

    number = {4},

}

@misc{nyt-large-type-1967,
  author       = {The New York Times},
  title        = {Times Begins an Edition Printed in Large Type},
  howpublished = {\url{https://www.nytimes.com/1967/03/06/archives/times-begins-an-edition-printed-in-large-type.html}},
  note         = {Accessed: August 14th, 2025},
  year         = {1967},
}

@article{reading-digital-legge,
author = {Legge, Gordon},
year = {2016},
month = {08},
pages = {102-125},
title = {Reading Digital with Low Vision},
volume = {50},
journal = {Visible language}
}

@article{magnification-low-vision,
author = {Christen, Michael and Abegg, Mathias},
year = {2017},
month = {05},
pages = {},
title = {The Effect of Magnification and Contrast on Reading Performance in Different Types of Simulated Low Vision},
volume = {10},
journal = {Journal of Eye Movement Research},
doi = {10.16910/jemr.10.2.5}
}

@article{bowers-reading,
author = {Bowers, Alex and Cheong, Allen and Lovie-Kitchin, Jan},
year = {2007},
month = {02},
pages = {9-20},
title = {Reading With Optical Magnifiers: Page Navigation Strategies and Difficulties},
volume = {84},
journal = {Optometry and vision science : official publication of the American Academy of Optometry},
doi = {10.1097/01.opx.0000254035.39055.05}
}

@article{responsive-design,
author = {Almeida, Fernando and Monteiro, J.},
year = {2017},
month = {12},
pages = {48-65},
title = {The role of responsive design in web development},
volume = {14},
journal = {Webology}
}

@misc{reflow,
author = {{W3C Web Accessibility Initiative (WAI)}},
title = {Understanding {SC} 1.4.10: Reflow (Level AA)},
howpublished = {\url{https://www.w3.org/WAI/WCAG21/Understanding/reflow.html}},
year = {2018},
note = {World Wide Web Consortium (W3C). Accessed 2025-08-13. Understanding document for WCAG~2.1, SC 1.4.10}
}

@article{holsanova_entry_2006,
	title = {Entry points and reading paths on newspaper spreads: comparing a semiotic analysis with eye-tracking measurements},
	volume = {5},
	issn = {1470-3572, 1741-3214},
	shorttitle = {Entry points and reading paths on newspaper spreads},
	url = {http://journals.sagepub.com/doi/10.1177/1470357206061005},
	doi = {10.1177/1470357206061005},
	language = {en},
	number = {1},
	urldate = {2023-06-30},
	journal = {Visual Communication},
	author = {Holsanova, Jana and Rahm, Henrik and Holmqvist, Kenneth},
	month = feb,
	year = {2006},
	pages = {65--93},
}

@TechReport{holmqvist_role_2005,
  author       = {Holmqvist, Kenneth and Wartenberg, Constanze},
  title        = {The role of local design factors for newspaper reading behaviour: An eye-tracking perspective},
  institution  = {Lund University, Cognitive Studies},
  type         = {LUCS Report 127},
  year         = {2005},
  address      = {Lund, Sweden},
  note         = {ISSN 1101-8453},
}

@Article{zambarbieri_eye_2008,
  author    = {Zambarbieri, Daniela and Carniglia, Elena and Robino, Carlo},
  title     = {Eye Tracking Analysis in Reading Online Newspapers},
  journal   = {Journal of Eye Movement Research},
  year      = {2008},
  volume    = {2},
  number    = {4},
  pages     = {7},
  doi       = {10.16910/jemr.2.4.7},
  url       = {https://bop.unibe.ch/JEMR/article/view/2279},
  month     = nov
}

@MastersThesis{holmberg_eye_2004,
  author       = {Holmberg, Nils},
  title        = {Eye movement patterns and newspaper design factors: An experimental approach},
  school       = {Lund University, Cognitive Science},
  year         = {2004},
  type         = {Master's Thesis},
  address      = {Lund, Sweden},
}

@Article{eraslan_eye_2015,
  author    = {Eraslan, Sukru and Yesilada, Yeliz and Harper, Simon},
  title     = {Eye tracking scanpath analysis techniques on web pages: A survey, evaluation and comparison},
  journal   = {Journal of Eye Movement Research},
  year      = {2015},
  volume    = {9},
  number    = {1},
  pages     = {2},
  doi       = {10.16910/jemr.9.1.2},
  url       = {https://www.mdpi.com/1995-8692/9/1/2},
}

@article{hart_nasa-task_2006,
author = {Sandra G. Hart},
title ={Nasa-Task Load Index (NASA-TLX); 20 Years Later},

journal = {Proceedings of the Human Factors and Ergonomics Society Annual Meeting},
volume = {50},
number = {9},
pages = {904-908},
year = {2006},
doi = {10.1177/154193120605000909},
}

@book{legge2006,
author = {Legge, Gordon E.},
address = {Mahwah, NJ},
booktitle = {Psychophysics of reading in normal and low vision},
isbn = {9780429175121},
language = {eng},
publisher = {Lawrence Erlbaum},
title = {Psychophysics of reading in normal and low vision / Gordon E. Legge and colleagues.},
year = {2006},
}

@article{beckmann_psychophysics_1996,
	title = {Psychophysics of {Reading}—{XIV}. {The} {Page} {Navigation} {Problem} in {Using} {Magnifiers}},
	volume = {36},
	copyright = {https://www.elsevier.com/tdm/userlicense/1.0/},
	issn = {00426989},
	url = {https://linkinghub.elsevier.com/retrieve/pii/0042698996000843},
	doi = {10.1016/0042-6989(96)00084-3},
	language = {en},
	number = {22},
	urldate = {2024-06-06},
	journal = {Vision Research},
	author = {Beckmann, Paul J. and Legge, Gordon E.},
	month = nov,
	year = {1996},
	pages = {3723--3733},
}

@article{xiong_digital_2022,
	title = {Digital {Reading} with {Low} {Vision}: {Principles} for {Selecting} {Display} {Size}},
	volume = {99},
	issn = {1538-9235, 1040-5488},
	shorttitle = {Digital {Reading} with {Low} {Vision}},
	url = {https://journals.lww.com/10.1097/OPX.0000000000001919},
	doi = {10.1097/OPX.0000000000001919},
	language = {en},
	number = {8},
	urldate = {2024-06-06},
	journal = {Optometry and Vision Science},
	author = {Xiong, Ying-Zi and Atilgan, Nilsu and Fletcher, Donald C. and Legge, Gordon E.},
	month = aug,
	year = {2022},
	pages = {655--661},
}

@article{calabrese_baseline_2016,
	title = {Baseline {MNREAD} {Measures} for {Normally} {Sighted} {Subjects} {From} {Childhood} to {Old} {Age}},
	volume = {57},
	copyright = {http://creativecommons.org/licenses/by-nc-nd/4.0/},
	issn = {1552-5783},
	url = {http://iovs.arvojournals.org/article.aspx?doi=10.1167/iovs.16-19580},
	doi = {10.1167/iovs.16-19580},
	language = {en},
	number = {8},
	urldate = {2025-05-16},
	journal = {Investigative Opthalmology \& Visual Science},
	author = {Calabrèse, Aurélie and Cheong, Allen M. Y. and Cheung, Sing-Hang and He, Yingchen and Kwon, MiYoung and Mansfield, J. Stephen and Subramanian, Ahalya and Yu, Deyue and Legge, Gordon E.},
	month = jul,
	year = {2016},
	pages = {3836},
}

@inproceedings{caine-2016,
author = {Caine, Kelly},
title = {Local Standards for Sample Size at CHI},
year = {2016},
isbn = {9781450333627},
publisher = {Association for Computing Machinery},
address = {New York, NY, USA},
url = {https://doi.org/10.1145/2858036.2858498},
doi = {10.1145/2858036.2858498},
booktitle = {Proceedings of the 2016 CHI Conference on Human Factors in Computing Systems},
pages = {981–992},
numpages = {12},
location = {San Jose, California, USA},
series = {CHI '16}
}

\appendix
\onecolumn

\section*{\added{Appendix: Linear-mixed effects model ANOVA tables}}
\label{appendix:anova}

\begin{table*}[ht]
    \centering
    
    \begin{tabular}{l c c c c c}
        \toprule
        \multicolumn{6}{c}{\textit{\textbf{Task 1 success ratio} - Fixed effects (type III tests)}} \\
        \midrule
        \textbf{Effect } & \textbf{Num DF} & \textbf{Den DF} & \textbf{Statistic} & \textbf{$p$-value} & \textbf{Effect Size} \\
        
        NewspaperID       & 5 & 215.94 & $F = 5.72$ & \textbf{<.001}  & $\eta_p^2 = 0.12$ \\
        \CVAS (Normal vs Low)       & 1 & 23.71 & $F = 20.92$ & \textbf{<.001} & $\eta_p^2 = 0.55$ \\
        Content Version (1,2,3)     & 2 & 223.45 & $F = 3.72$  & \textbf{0.025}  & $\eta_p^2 = 0.03$ \\
        NewspaperID $\times$ CVAS     & 5 & 216.1 & $F = 3.83$  & \textbf{.002}  & $\eta_p^2 = 0.08$ \\
        
        \bottomrule
    \end{tabular}
        \caption{\label{tab:anova-success-ratio} Task 1 success ratio}
        \Description{A statistical table reporting significant results for Task 1 success ratio. The rows list the effects for NewspaperID, CVAS, Content Version and interaction of NewspaperID $\times$ CVAS. The columns include Degrees of Freedom, F-statistic, p-value, and Effect Size.}
\end{table*}



\begin{table*}[h]
    \centering

    \begin{tabular}{l c c c c c}
        \toprule
        \multicolumn{6}{c}{\textit{\textbf{Task 1 completion time} - Fixed effects (type III tests)}} \\
        \midrule
        \textbf{Effect } & \textbf{Num DF} & \textbf{Den DF} & \textbf{Statistic} & \textbf{$p$-value} & \textbf{Effect Size} \\
        Condition (\condonenot vs \condtwonot)      & 1 & 209.43 & $F = 116.5$ & \textbf{<.001}  & $\eta_p^2 = 0.42$ \\
        
        NewspaperID       & 5 & 209.89 & $F = 9.57$ & \textbf{<.001}  & $\eta_p^2 = 0.19$ \\
        \CVAS (Normal vs Low)       & 1 & 18.61 & $F = 50.19$ & \textbf{<.001} & $\eta_p^2 = 0.73$ \\
        Content Version (1,2,3)     & 2 & 213.5 & $F = 3.21$  & \textbf{0.04}  & $\eta_p^2 = 0.03$ \\
        Condition $\times$ CVAS     & 1 & 209.18 & $F = 146.3$  & \textbf{<.001}  & $\eta_p^2 = 0.41$ \\
        
        \bottomrule
    \end{tabular}

        \caption{Task 1 completion time}
        \Description{A statistical table reporting significant results for Task 1 completion time. The rows list the effects for Condition, NewspaperID, CVAS, Content Version, and the interaction of Condition $\times$ CVAS. The columns include Degrees of Freedom, F-statistic, p-value, and Effect Size.}
\end{table*}



\begin{table*}[h]
    \centering

    \begin{tabular}{l c c c c c}
        \toprule
        \multicolumn{6}{c}{\textit{\textbf{Task 1 reading path similarity} - Fixed effects (type III tests)}} \\
        \midrule
        \textbf{Effect } & \textbf{Num DF} & \textbf{Den DF} & \textbf{Statistic} & \textbf{$p$-value} & \textbf{Effect Size} \\
        Condition (\condonenot vs \condtwonot)      & 1 & 210.75 & $F = 11.98$ & \textbf{<.001}  & $\eta_p^2 = 0.14$ \\
        
        NewspaperID       & 5 & 212.51 & $F = 6.15$ & \textbf{<.001}  & $\eta_p^2 = 0.13$ \\
        Reading frequency     & 3 & 212.49 & $F = 5.94$  & \textbf{<.001}  & $\eta_p^2 = 0.08$ \\

        \midrule
        
        \multicolumn{6}{c}{\textit{\textbf{Post-hoc comparison}}} \\
        
        NewspaperID (Layout \#19 vs others)      & 5 & 209.19 & $F = 8.93$ & \textbf{<.001}  & $\eta_p^2 = 0.18$ \\
        
        \bottomrule
    \end{tabular}

        \caption{Task 1 reading path similarity}
        \Description{A statistical table reporting significant results for Task 1 reading path similarity. The rows list the effects for Condition, NewspaperID, and Reading frequency. The columns include Degrees of Freedom, F-statistic, p-value, and Effect Size.}
\end{table*}

\begin{table*}[h]
    \centering

    \begin{tabular}{l c c c c c}
        \toprule
        \multicolumn{6}{c}{\textit{\textbf{Task 1 reading time} - Fixed effects (type III tests)}} \\
        \midrule
        \textbf{Effect } & \textbf{Num DF} & \textbf{Den DF} & \textbf{Statistic} & \textbf{$p$-value} & \textbf{Effect Size} \\
        Condition (\condonenot vs \condtwonot)      & 1 & 209.24 & $F = 6.02$ & \textbf{0.01}  & $\eta_p^2 = 0.05$ \\
        
        \CVAS (Normal vs Low)       & 1 & 18.3 & $F = 15.5$ & \textbf{<.001}  & $\eta_p^2 = 0.46$ \\
        Trial Index (1...6)    & 1 & 209.13 & $F = 7.66$  & \textbf{.006}  & $\eta_p^2 = 0.04$ \\
        Content Version (1,2,3)     & 2 & 211.57 & $F = 3.72$  & \textbf{0.02}  & $\eta_p^2 = 0.03$ \\
         Condition $\times$ NewspaperID     & 5 & 209.02 & $F = 4.97$  & \textbf{<.001}  & $\eta_p^2 = 0.11$ \\

        \midrule

        \multicolumn{6}{c}{\textit{\textbf{Task 1 transition time} - Fixed effects (type III tests)}} \\
        \midrule
        \textbf{Effect } & \textbf{Num DF} & \textbf{Den DF} & \textbf{Statistic} & \textbf{$p$-value} & \textbf{Effect Size} \\
        Condition (\condonenot vs \condtwonot)      & 1 & 210.09 & $F = 73.57$ & \textbf{<.001}  & $\eta_p^2 = 0.29$ \\
        
        \CVAS (Normal vs Low)       & 1 & 19.58 & $F = 71.6$ & \textbf{<.001}  & $\eta_p^2 = 0.8$ \\
        NewspaperID     & 5 & 211.23 & $F = 7.45$  & \textbf{<.001}  & $\eta_p^2 = 0.15$ \\
        Trial Index (1...6)    & 1 & 209.62 & $F = 4.38$  & \textbf{.037}  & $\eta_p^2 = 0.02$ \\
        
       Condition $\times$ \CVAS     & 1 & 209.47 & $F = 113.82$  & \textbf{<.001}  & $\eta_p^2 = 0.35$ \\
    
        \bottomrule
    \end{tabular}
    
        \caption{Task 1 reading and transition time}
        \Description{A statistical table reporting significant results for Task 1 reading and transition time. The rows list the effects for both cases: (1) In the case of Task 1 reading time, the rows are Condition, CVAS, Trial index, Content version, and the interaction Condition $\times$ NewspaperID. (2)  In the case of Task 1 transition time, the rows are Condition, CVAS, NewspaperID, Trial index, Content version, and the interaction Condition $\times$ CVAS. The columns include Degrees of Freedom, F-statistic, p-value, and Effect Size.}

\ \\[1cm]
    \centering

    \begin{tabular}{l c c c c c}
        \toprule
        \multicolumn{6}{c}{\textit{\textbf{Task 2 success ratio} - Fixed effects (type III tests)}} \\
        \midrule
        \textbf{Effect } & \textbf{Num DF} & \textbf{Den DF} & \textbf{Statistic} & \textbf{$p$-value} & \textbf{Effect Size} \\
        NewspaperID       & 5 & 219.47 & $F = 4.10$ & \textbf{.001}  & $\eta_p^2 = 0.09$ \\
        NewspaperID $\times$ CVAS     & 5 & 219.66 & $F = 3.30$  & \textbf{.006}  & $\eta_p^2 = 0.07$ \\
        
        \bottomrule
    \end{tabular}

        \caption{Task 2 success ratio}
        \Description{A statistical table reporting significant results for Task 2 success ratio. The rows list the effects for NewspaperID, and the interaction between NewspaperID $\times$ CVAS. The columns include Degrees of Freedom, F-statistic, p-value, and Effect Size.}

\ \\[1cm]

    \centering

    \begin{tabular}{l c c c c c}
        \toprule
        \multicolumn{6}{c}{\textit{\textbf{Task 2 completion time} - Fixed effects (type III tests)}} \\
        \midrule
        \textbf{Effect } & \textbf{Num DF} & \textbf{Den DF} & \textbf{Statistic} & \textbf{$p$-value} & \textbf{Effect Size} \\
        Condition      & 1 & 212.63 & $F = 15.06$ & \textbf{<.001}  & $\eta_p^2 = 0.12$ \\
        \CVAS (Normal vs Low)   & 1 & 23.55 & $F = 47.20$  & \textbf{<.001}  & $\eta_p^2 = 0.71$ \\
        Condition $\times$ \CVAS      & 1 & 210.63 & $F = 14.81$  & \textbf{<.001}  & $\eta_p^2 = 0.07$ \\
        \bottomrule
    \end{tabular}

        \caption{Task 2 completion time}
        \Description{A statistical table reporting significant results for Task 2 completion time. The rows list the effects for Condition, CVAS, and the interaction between Condition $\times$ CVAS. The columns include Degrees of Freedom, F-statistic, p-value, and Effect Size.}
\end{table*}

\end{document}